\documentclass{aa}  
\usepackage{graphicx}
\usepackage{natbib}

\usepackage{amsmath,amssymb,amsfonts,bm}
\usepackage{hyperref}
\usepackage{calc}
\usepackage{color}
\usepackage{wrapfig}
\usepackage{url}
\usepackage{paralist}
\usepackage{array}
\usepackage{tabularx}
\usepackage{txfonts}
\DeclareMathOperator \arcsinh {arsinh}

\begin{document}

\bibliographystyle{aa/aa}

   \title{Asymmetric Gravitational Lenses in TeVeS\\and Application to the Bullet Cluster}


   \author{M. Feix
          \and
          C. Fedeli
          \and
          M. Bartelmann 
          }

   \institute{Zentrum f\"ur Astronomie, ITA, Universit\"at Heidelberg,
              Albert-\"Uberle-Str. 2, D-69120 Heidelberg\\
              \email{mfeix@ita.uni-heidelberg.de}
             }


 
  \abstract
   {}
   {We explore the lensing properties of asymmetric matter density distributions in Bekenstein's Tensor-Vector-Scalar theory (TeVeS).}
   {Using an iterative Fourier-based solver for the resulting non-linear scalar field equation, we numerically calculate the total gravitational potential and derive the corresponding TeVeS lensing maps.}
   {Considering variations on rather small scales, we show that the lensing properties significantly depend on the lens's extent along the line of sight. Furthermore, all simulated TeVeS convergence maps strongly track the dominant baryonic components, non-linear effects, being capable of counteracting this trend, turn out to be very small. Setting up a toy model for the cluster merger $1$E$0657-558$, we infer that TeVeS cannot explain observations without assuming an additional dark mass component in both cluster centers, which is in accordance with previous work.}
   {}

   \keywords{MOND --
                TeVeS --
                Gravitational Lensing
               }

   \maketitle
%

\section{Introduction}

As is known, General Relativity (GR) cannot explain the dynamics of our universe on large physical scales as the amount of visible mass clearly lies below what would be expected from the observed gravitational effects. Commonly, this is denoted as the Missing Mass Problem. The usual remedy is to invoke a form of matter which does not couple to light, therefore being referred to as Dark Matter (DM). Over the last years, this paradigm has been remarkably successful in forming a consistent cosmological picture since it provides suitable explanations for the observations on supernovae Ia \citep{sne}, large-scale structure \citep{largescale,2dFGRS}, and the CMB \citep{cmb}.

However, one can also take a different point of view and modify the law of gravity itself. In the past, there have been several suggestions for such modifications: $f(R)$ gravity \citep{fr}, conformal Weyl gravity and Aether-type theories \citep{tv2}, to name just a few. A special realization of the latter, the so-called Tensor-Vector-Scalar gravity (TeVeS) \citep{teves,tv1} has recently gained interest as it provides a fully relativistic framework for the Modified Newtonian Dynamics (MOND) paradigm { \citep{Mond3, mondnew}}. Compared to other modifications, MONDian dynamics is characterized by an acceleration scale $a_{0}$, and its departure from classical Newtonian predictions depends on acceleration:
\begin{equation}
\tilde{\mu}\left(\frac{|\vec{a}|}{a_{0}}\right)\vec{a} = - \vec\nabla\Phi_{N} +{\vec S}.
\label{eq:1}
\end{equation}
Here, $\Phi_{N}$ denotes the common Newtonian potential of the visible (baryonic) matter {and $\vec S$ is a solenoidal vector field determined by the condition that $\vec a$ can be expressed as the gradient of a scalar potential. The function $\tilde\mu$, controlling the modification of Newton's law, has the following asymptotic behavior:}
\begin{equation}
\begin{split}
\tilde\mu(x) \sim x  \qquad x \ll 1,\\
\tilde\mu(x) \sim 1  \qquad x \gg 1.
\end{split}
\label{eq:2}
\end{equation}
Eq. \eqref{eq:1} has been constructed to agree with the fact that the rotation curves of spiral galaxies become flat outside their central parts { \citep{spiral1}}. Analyzing observational data, Milgrom estimated $a_{0} \approx 1 \times 10^{-10}$m/s$^{2}$. Within this work, we shall study TeVeS and its built-in MONDian dynamics in the context of gravitational lensing, focusing on non-spherical density distributions.

The paper is structured as follows: Starting with a brief introduction to TeVeS and the formalism of gravitational lensing, we shall have a look at analytic lens models in spherical symmetry. In TeVeS, the effect of gravity on matter is controlled by a function which is only constrained by its non-relativistic limits, i.e. the Newtonian and the MONDian limit, with the function's actual form in the intermediate part providing an additional degree of freedom to the theory. Therefore, we shall proceed with a general analysis of this so-called free function's influence on the TeVeS deflection angle.
Choosing a specific form of the free function, we will present a numerical tool that allows the treatment of non-spherical lenses in TeVeS. In contrast to already existent MOND solvers \citep{numericmond,multigrid,nbody} our method is based on fast Fourier techniques, achieving high-resolution solutions for the TeVeS scalar potential { on time scales up to a few hours on standard PCs}. Highlighting basic properties and limitations of this method, we shall apply it to a set of different matter density distributions including discussions on the failure of the thin lens approximation and general properties of the TeVeS lensing maps. Finally, we will create a toy model for the cluster merger $1$E$0657-558$, with the resulting lensing maps confirming and extending the basic findings of \cite{tevesfit}.


\section{Fundamentals of TeVeS}
In the following, we will give a brief review on TeVeS and the approximations used for quasi-static systems like galaxies (or galaxy clusters) and cosmology. { If not specified in any other way, the ``square" of a vector denotes the square of its Euclidean norm, i.e. $(\vec A)^{2}=(\lVert\vec A\rVert_{2})^2$.} Henceforth, we shall use units with $c=1$.

\subsection{Fields and Actions}
\label{section21}
TeVeS gravity is based on three dynamical fields: an Einstein metric $g_{\mu\nu}$, a vector field $U_{\mu}$ such that
\begin{equation}
g^{\mu\nu}U_{\mu}U_{\nu} = -1
\label{eq:3}
\end{equation}
and a scalar field $\phi$. An essential feature of TeVeS is the introduction of a physical frame described by the metric $\tilde{g}_{\mu\nu}$ which is needed for gravity-matter coupling only and obtained from the non-conformal relation
\begin{equation}
\tilde g_{\mu\nu} = e^{-2\phi}g_{\mu\nu}-2U_{\mu}U_{\nu}\sinh(2\phi).
\label{eq:4}
\end{equation}
The geometrical part of the action is exactly the same as in GR:
\begin{equation}
S_{g} = {\frac{1}{16\pi G}}\int g^{\mu\nu}R_{\mu\nu}\sqrt{-g}d^{4}x,
\label{eq:5}
\end{equation}
where $R_{\mu\nu}$ is the Ricci tensor of $g_{\mu\nu}$ and $g$ the determinant of $g_{\mu\nu}$. Note that the TeVeS constant $G$ must not be mistaken for the Newtonian gravitational constant $G_{N}$ (cf. Sec. \ref{section23}). The vector field's action $S_{v}$ reads as follows:
\begin{equation}
S_{v} = -\frac{K}{32\pi G}\int\left\lbrack F^{\mu\nu}F_{\mu\nu}-\lambda(g^{\mu\nu}U_{\mu}U_{\nu}+1)\right\rbrack\sqrt{-g}d^{4}x,
\label{eq:6}
\end{equation}
with $F_{\mu\nu} = U_{\mu,\nu}-U_{\nu,\mu}$. Here the constant $K$ describes the vector's coupling to gravity and $\lambda$ is a Lagrangian multiplier enforcing the normalization given by Eq. \eqref{eq:3}. Eq. \eqref{eq:6} corresponds to the classical Maxwell action, the field $U_{\mu}$ now having an effective mass. The action $S_{s}$ of the scalar field $\phi$ involves an additional non-dynamical scalar field $\sigma$, and takes the form
\begin{equation}
S_{s} = -\frac{1}{2}\int\left\lbrack\sigma^{2}h^{\mu\nu}\phi_{,\mu}\phi_{,\nu}+\frac{G\sigma^{4}}{2l^{2}}F(kG\sigma^{2})\right\rbrack\sqrt{-g}d^{4}x,
\label{eq:7}
\end{equation}
where $h^{\mu\nu} = g^{\mu\nu}-U^{\mu}U^{\nu}$ and $F$ is a dimensionless free function. As the field $\sigma$ is related to the invariant $h^{\mu\nu}\phi_{,\mu}\phi_{,\nu}$, however, it could in principle be eliminated from the action. While $k$ is the coupling constant of $\phi$ to gravity, the constant $l$ is related to Milgrom's $a_{0}$ and has the dimension of a length (see Sec. \ref{section22}).
Finally, according to the equivalence principle, the matter action is given by
\begin{equation}
S_{m} = \int\mathcal{L}_{m}\sqrt{-\tilde{g}}d^{4}x.
\label{eq:8}
\end{equation}
Matter fields are coupled to gravity by the physical metric $\tilde{g}_{\mu\nu}$, i.e. world lines are geodesics of the metric $\tilde{g}_{\mu\nu}$ rather than ${g}_{\mu\nu}$. As usual, the corresponding equations of motion can be derived by varying the total action $S=S_{g}+S_{v}+S_{s}+S_{m}$ w.r.t. the basic fields.

In order to obtain Newton's law in the non-relativistic high acceleration regime $(a \gg a_{0})$, the coupling constants $k$ and $K$ have to be small, i.e.
\begin{equation}
k \ll 1,\quad K \ll 1.
\label{eq:9}
\end{equation}
Therefore, TeVeS is kept close to GR in a sense that it will recover well-known features of GR, albeit modified by the other fields.
\subsection{The Free Function}
\label{section22}
In TeVeS, the transition from Newtonian dynamics to MOND is controlled by the free function $F$. Following \cite{teves},
the ``equation of motion" for the non-dynamical field $\sigma$ suggests introducing a new function $\mu(y)$ which is implicitly given by
\begin{equation}
-\mu F(\mu)-\frac{1}{2}\mu^{2}F^{'}(\mu) = y,
\label{eq:10}
\end{equation}
with
\begin{equation}
kG\sigma^{2} = \mu(kl^{2}h^{\mu\nu}\phi_{,\mu}\phi_{,\nu}) = \mu(y).
\label{eq:11}
\end{equation}
For further analysis, we shall assume the function $\mu(y)$ to behave well in a physical sense, i.e to be smooth and monotonic in both cosmological $(y<0)$ and quasi-static situations $(y>0)$.
In order to reproduce both a MONDian and a Newtonian limit, the quasi-static branch of the inverse function $y(\mu)$ has to satisfy the following conditions:
\begin{equation}
\begin{split}
y(\mu) \rightarrow \infty,\quad\mu\rightarrow 1,\\
{
y(\mu) \sim b\mu^{2},\quad\mu\ll 1,
}
\label{eq:12}
\end{split}
\end{equation}
where $b$ is a positive real constant. If this is the case, the constant $l$ can be related to Milgrom's $a_{0}$ by
\begin{equation}
a_{0} = \frac{\sqrt{bk}}{4\pi\Xi l} \approx \frac{\sqrt{bk}}{4\pi l},
\label{eq:13}
\end{equation}
where $\Xi=1-K/2-2\phi_{c}$ and $\phi_{c}$ is the cosmological value of the scalar field which is assumed to be small ($\phi_{c} \ll 1$).
In Sec. \ref{parameter}, we shall return to the free function and its properties in the context of gravitational lensing, concentrating on the branch relevant for quasi-static systems.
\subsection{Quasi-static Systems}
\label{section23}
According to \cite{teves}, the physical metric field near a quasi-static galaxy (cluster) is identical to the metric obtained in GR if the non-relativistic gravitational potential is replaced by
\begin{equation}
\Phi = \Xi\Phi_{N}+\phi,
\label{eq:14}
\end{equation}
where $\Phi_{N}$ is the Newtonian potential generated by the baryonic matter density $\rho$. In this approximation, it is consistent to take $U^{\mu} = (U^{t},0,0,0)$ which can be shown from the corresponding field equations. Then we have
\begin{equation}
kl^{2}h^{\mu\nu}\phi_{,\mu}\phi_{,\nu} \rightarrow kl^{2}(\vec\nabla\phi)^{2}
\end{equation}
and the equation of the scalar field reduces to
\begin{equation}
\vec\nabla\left\lbrack\mu\left(kl^{2}(\vec\nabla\phi)^{2}\right)\vec\nabla\phi\right\rbrack = kG\rho.
\label{eq:15}
\end{equation}
Eq. \eqref{eq:15} corresponds to the non-linear elliptic boundary value problem and can be treated numerically. In Sec. \ref{sectionsolver}, we shall give a detailed description of the method we use to determine the solution for the scalar field $\phi$, including a discussion on its problems and limitations.

Since we have $K,\phi_{c}\ll 1$, the quantity $\Xi$ has a value close to unity, and the total potential $\Phi$ can essentially be written as the sum of the common Newtonian potential $\Phi_{N}$ and the additional scalar field, i.e. Eq. \eqref{eq:14} may further be reduced to
\begin{equation}
\Phi = \Phi_{N}+\phi.
\label{eq:16}
\end{equation}
As the constant $G$ is related to the Newtonian gravitational constant $G_{N}$ by \citep{teves}
\begin{equation}
G_{N} = \left(\Xi+\frac{k}{4\pi}\right)G,
\label{constantg}
\end{equation}
we will additionally assume $G\approx\ G_{N}$ throughout this work.

\subsection{Cosmology}
\label{section24}
Similar to the case of GR, it is possible to derive a cosmological model in TeVeS. Assuming the basic fields to partake of the symmetries of the Friedmann-Robertson-Walker (FRW) spacetime, the analog of Friedmann's equation reads
\begin{equation}
\left(\frac{\dot a}{a}\right)^{2} = \frac{8\pi G}{3}(\rho e^{-2\phi}+\rho_{\phi})-\frac{K}{a^{2}}+\frac{\Lambda}{3},
\label{eq:17}
\end{equation}
where $\rho_{\phi}$ is the energy density of the scalar field given by
\begin{align}
\rho_{\phi} = \frac{\mu\dot\phi^{2}}{kG}+\frac{\mu^{2}}{4k^{2}l^{2}G}F(\mu) = \frac{-2\mu y(\mu)+\mu^{2}F(\mu)}{4k^{2}l^{2}G}.
\label{eq:18}
\end{align}
Since we are interested in the physical metric, we have to make use of transformation \eqref{eq:4} and finally obtain
\begin{equation}
\frac{1}{\tilde{a}}\frac{d\tilde{a}}{d\tilde{t}} = e^{-\phi}\left(\frac{\dot{a}}{a}-\dot\phi\right),
\label{eq:19}
\end{equation}
with
\begin{equation}
d\tilde{t} = e^{\phi}dt,\quad \tilde{a} = e^{-\phi}a.
\label{eq:20}
\end{equation}
In order to simplify matters, however, we shall introduce the ``minimal" cosmological model proposed by \cite{lenstest}:

According to \cite{teves}, it is consistent to assume that the cosmological scalar field evolves slowly in time throughout cosmological history. Thus, its contribution to the Hubble expansion is negligibly small, with a ratio $\mathcal{O}(k)$ compared to the matter contribution. Setting $\rho_{\phi}=0$ and recalling that $\phi\ll 1$, the physical Hubble parameter can be expressed as
\begin{equation}
\tilde{H}^{2} \approx H^{2} \approx H_{0}^{2}\left(\Omega_{m}(1+z)^{3}+\Omega_{\Lambda}+\Omega_{K}(1+z)^{2}\right),
\label{eq:21}
\end{equation}
where $\Omega_{K} \approx 1-\Omega_{\Lambda}-\Omega_{m}$.
Since there is no DM in TeVeS, we have to consider a minimal-matter cosmology that should be consistent with observational data in order to obtain a reasonable cosmological model. \citeauthor{lenstest} actually find a good fit of the high-z SNe distance moduli data set by choosing an open cosmology with $\Omega_{\Lambda} \sim 0.46$, $\Omega_{m} \sim 0.04$ and $H_{0} \sim 70$km/s/Mpc. However, they also point out that when moving to very high redshifts, this open cosmology has problems, i.e. it underestimates the last scattering sound horizon, which actually seems to be an artifact of the crude approximation as recent work has shown \citep{tevescosmo}. Nevertheless, in the context of gravitational lensing, this simple model is sufficient for assigning the distances of lenses and sources up to a redshift of $z \sim 3$.

\section{Gravitational Lensing in TeVeS}
\subsection{Light Bending in Slightly Curved Spacetime}
\label{section31}
In general, light rays move along the null geodesics of the underlying metric field, i.e. the null geodesics of the physical metric $\tilde{g}_{\mu\nu}$ considering the framework of TeVeS gravity. For an on average homogeneous and isotropic universe with local perturbations, however, light mostly travels through unperturbed spacetime and is only deflected close to inhomogeneities which act as gravitational lenses. If the non-relativistic potential $\Phi$ and the peculiar velocity $v$ of the lens are small ($\Phi,v \ll 1$), we can presume a locally flat spacetime being disturbed by the potential $\Phi$; these conditions are well satisfied for galaxies and galaxy clusters.

It is well-known \citep{gl} that the deflection angle of a light ray under these assumptions can be expressed as
\begin{equation}
\vec{\hat{\alpha}} = 2\int_{-\infty}^{\infty}\vec\nabla_{\bot}\Phi dl = \vec{\hat{\alpha}}_{GR}+2\int_{-\infty}^{\infty}\vec\nabla_{\bot}\phi dl,
\label{eq:22}
\end{equation}
where $\Phi$ is given by Eq. \eqref{eq:16}, { $\vec\nabla_{\bot}$ denotes the two-dimensional gradient operator perpendicular to light propagation} and integration is performed along the unperturbed light path (Born's approximation). In addition to the deflection angle caused by the Newtonian potential $\Phi_{N}$, there is a contribution arising from the scalar field $\phi$. Because $\phi$ is connected to the matter density in a highly non-linear way, it is not possible to relate the projected matter density to a two-dimensional scalar deflection potential just like in GR (cf. Sec. \ref{section32}). Therefore, we are obliged to solve Eq. \eqref{eq:15} for calculating the TeVeS deflection angle, which is a very delicate issue (cf. Sec. \ref{sectionsolver}). Compared to the distances between lens and source and observer and source, however, we may still assume that most of the bending occurs within a small range around the lens. This enables us to fully adopt the GR lensing formalism which will be introduced in the next section. For further discussion, we shall choose coordinates such that unperturbed light rays propagate parallel to the $z$-axis.

\subsection{Lensing Formalism and Critical Curves}
\label{section32}
In gravitational lensing, it is convenient to introduce the deflection potential $\Psi(\vec\theta)$ \citep{gl}:
\begin{equation}
\Psi(\vec\theta) = 2\frac{D_{ds}}{D_{s}D_{d}}\int\Phi(D_{d}\vec\theta,z)dz,
\label{eq:23}
\end{equation}
where we have used $\vec\theta=\vec\xi/D_{d}$. Here $\vec\xi$ is the 2-dimensional position vector in the lens plane, and $D_{s}$, $D_{d}$, and $D_{ds}$ are the (angular diameter) distances between source and observer, lens and observer, and lens and source, respectively. If a source is much smaller than the angular scale on which the lens properties change, the lens mapping can locally be linearized. Thus, the distortion of an image can be described by the Jacobian matrix
\begin{equation}
\mathcal{A}(\vec\theta) = \frac{\partial\vec\beta}{\partial\vec\theta} =
\begin{pmatrix}
1-\kappa-\gamma_{1} & -\gamma_{2}\\
-\gamma_{2} & 1-\kappa+\gamma_{1}
\end{pmatrix},
\label{eq:24}
\end{equation}
where $\vec\beta$=$\vec\eta/D_{s}$ and $\vec\eta$ denotes the 2-dimensional position of the source. The convergence $\kappa$ is directly related to the deflection potential $\Psi$ through
\begin{equation}
\kappa = \frac{1}{2}\Delta_{\vec\theta}\Psi
\label{eq:25}
\end{equation}
and the shear components $\gamma_{1},\gamma_{2}$ are given by
\begin{equation}
\gamma_{1} = \frac{1}{2}\left(\frac{\partial^{2}\Psi}{\partial\theta_{1}^{2}}-\frac{\partial^{2}\Psi}{\partial\theta_{2}^{2}}\right),\quad\gamma_{2} = \frac{\partial^{2}\Psi}{\partial\theta_{1}\partial\theta_{2}}, \quad\gamma=\sqrt{\gamma_{1}^{2}+\gamma_{2}^{2}}.
\label{eq:26}
\end{equation}
Points in the lens plane where
\begin{equation}
\det\mathcal{A}=(1-\kappa-\gamma)(1-\kappa+\gamma)=0,
\label{eq:27}
\end{equation}
form closed curves, the critical curves. Their image curves located in the source plane are called caustics. Since images near critical curves can significantly be magnified and distorted, which, for instance, is indicated by the giant luminous arcs formed from source galaxies near caustics, a thorough analysis of the behavior of such curves in a TeVeS universe will be profitable.

\subsection{Analytic Solutions}
\label{section33}
\begin{figure}
   \centering
   \includegraphics[width=\linewidth]{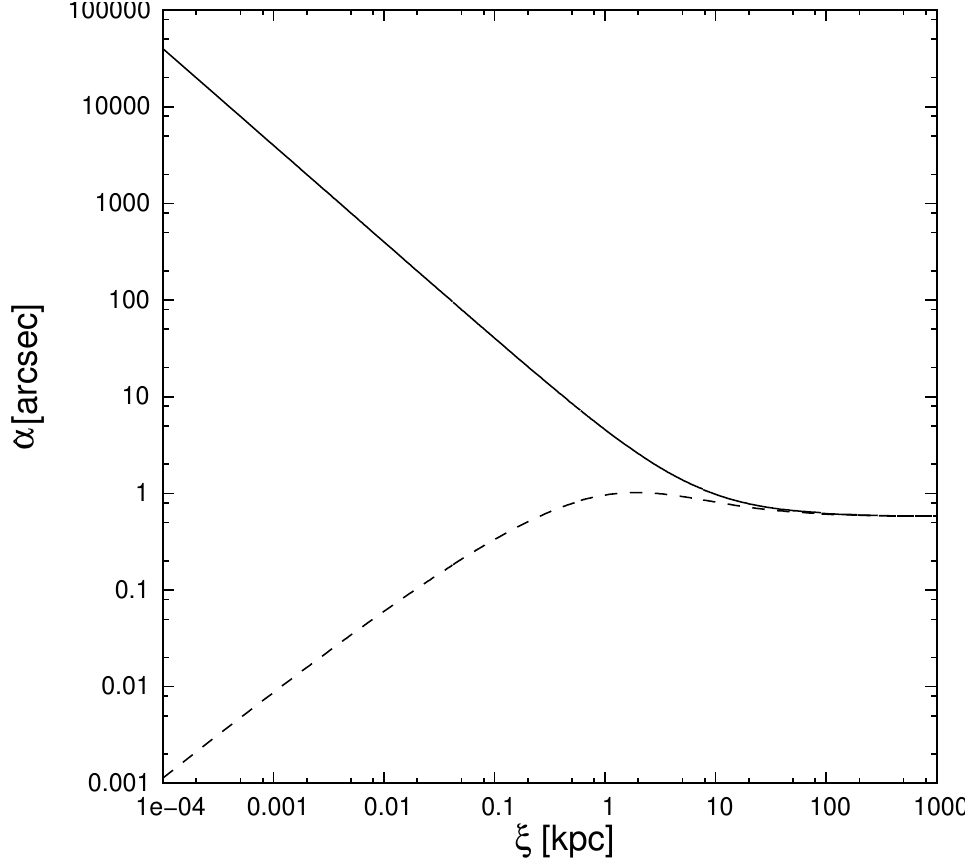}
\caption{The TeVeS Hernquist lens (dashed) with $r_{H}=2$kpc compared to its corresponding point lens (solid) w.r.t. the deflection angle $\hat\alpha$ for $M=10^{11}M_{\odot}$, $a_{0} = 1 \times 10^{-10}$m/s$^{2}$ and $D = D_{ds}D_{d}/D_{s} = 850$Mpc. For $\xi\gg\ 1$, both angles approach the constant $\hat\alpha_{\infty}=2\pi\sqrt{GMa_{0}}\approx 0.58^{''}$. The transition to the MONDian regime can be characterized by the critical radius $r_{0} = \sqrt{GM/a_{0}} \approx 10$kpc.
}
         \label{hernpoint}
   \end{figure}
Following \cite{lenstest}, we switch to a notation which turns out to be more suitable for analytic studies. Instead of the function $\mu$, we shall consider a new function $\bar{\mu}$ which is defined by
\begin{equation}
\frac{\bar\mu}{1-\bar\mu} = \frac{4\pi}{k}\left(1-\frac{K}{2}\right)^{-1}\mu,
\label{eq:28}
\end{equation}
where $k$, $K$ are the coupling constants of the scalar field $\phi$ and the vector field $U_{\mu}$, respectively. Similarly, we can relate the function $y$ to another function $\delta_{\phi}$ in the following way:
\begin{equation}
\delta_{\phi}^{2} = \left(\frac{4\pi}{k}\left(1-\frac{K}{2}\right)\right)^{2}\frac{y}{b} \approx \frac{(\vec\nabla\phi)^{2}}{a_{0}^{2}},
\label{eq:29}
\end{equation}
where $b$ is the real-valued parameter of the function $y(\mu)$ in \eqref{eq:12}. Choosing the free function such that
\begin{equation}
\delta_{\phi}^{2} \approx \frac{\bar\mu^{2}}{(1-\bar\mu)^{2}},\quad \bar\mu^{2} \approx \frac{\delta_{\phi}^{2}}{(1+\delta_{\phi})^{2}},
\label{eq:30}
\end{equation}
it is possible to obtain an analytic expression for the deflection angle of a Hernquist lens, i.e. a lens whose matter distribution follows a Hernquist profile \citep{hernquist} given by
\begin{equation}
\rho(r) = \frac{Mr_{H}}{2\pi r(r+r_{H})^{3}},
\label{eq:31}
\end{equation}
the Hernquist radius $r_{H}$ being the core scale length and $M$ the total mass. Eq. \eqref{eq:31} is a spherical profile which closely approximates the de Vaucouleurs $R^{1/4}$-law for elliptical galaxies. Using elementary calculus, we eventually end up with
\begin{equation}
\hat\alpha(\xi) = \frac{r_{H}A(\xi)}{\sqrt{|\xi^{2}-r_{H}^{2}|}}\left(4\xi\sqrt{GMa_{0}}+\frac{4GM\xi}{|\xi^{2}-r_{H}^{2}|}\right)-\frac{4GM\xi}{|\xi^{2}-r_{H}^{2}|},
\label{eq:32}
\end{equation}
where
\begin{equation}
A(\xi) = 
\begin{cases}
\arcsinh{\sqrt{\left|1-\left(r_{H}/\xi\right)^{2}\right|}} & \xi<r_{H}\\
\arcsin{\sqrt{1-\left(r_{H}/\xi\right)^{2}}} & \xi>r_{H}
\end{cases}.
\label{eq:33}
\end{equation}
In the limit $r_{H}\rightarrow 0$, the Hernquist lens coincides with a point lens. In this case, we find that the deflection angle is given by
\begin{equation}
\hat{\alpha}(\xi) = \frac{4GM}{\xi}+2\pi\sqrt{GMa_{0}}.
\label{eq:34}
\end{equation}
Obviously, the scalar part of TeVeS gravity of a point mass seems to mimic the presence of a dark isothermal sphere. Therefore, both GR including DM and TeVeS will essentially make the same lensing predictions for $\xi$ being much larger than the extension of the lens, but the highly non-linear coupling of the scalar field strongly suggests that there may be significant differences when moving to the strong acceleration regime near the center. Note that, although Eqs. \eqref{eq:32} and \eqref{eq:34} do not explicitly depend on $k$ and $K$, $a_{0}$ is still given by Eq. \eqref{eq:13}.

Fig. \ref{hernpoint} shows the lensing properties of both the TeVeS Hernquist lens ($r_{H}=2$kpc) and its corresponding point lens where we have set $M=10^{11}M_{\odot}$, $a_{0} = \sqrt{bk}/4\pi l =1 \times 10^{-10}$m/s$^{2}$ and $D = D_{ds}D_{d}/D_{s} = 850$Mpc. Since its deflection angle can be expressed analytically, the Hernquist lens is a perfectly suitable candidate for testing an algorithm for non-spherical problems.

\section{Influence of the Free Function}
\label{spherical}
Considering a spherically symmetric situation { and applying Gauss's theorem for a spherical surface of arbitrary radius}, Eq. \eqref{eq:15} can be transformed into
\begin{equation}
\vec\nabla\phi = \frac{k}{4\pi\mu}\vec\nabla\Phi_{N}.
\label{eq:35}
\end{equation}
Assuming we already know $\Phi_{N}$, for example by solving Poisson's equation, the relation above can directly be used to calculate $\vec\nabla\phi$ for any given function $\mu(y)$ (Remember that $y=kl^{2}(\vec\nabla\phi)^{2}$ for quasi-static systems, with $l$ given by Eq. \eqref{eq:13}). If $\mu$ or $\Phi_{N}$ cannot be obtained analytically, treatment with numerical methods, which can easily be applied in the spherically symmetric case, becomes necessary.
Because of their simplicity, spherically symmetric systems are particularly suitable for investigating the effects of the free function $y(\mu)$ on the deflection angle.

\subsection{Parameterization of the Free Function}
\label{parameter}
Having set the cosmological background in Sec. \ref{section24}, we shall focus on the free function's quasi-static branch $(y>0)$: If $y(\mu)$ can be analytically continued into the ring domain $R = \lbrace z \in \mathbb{C} \mid 0<|z-1|<1\rbrace$, it can be expanded into a Laurent series. Thus, $y(\mu)$ takes the following form for $0<\mu<1$:
\begin{equation}
y(\mu) = \sum_{n=1}^{\infty}\frac{a_{n}}{(1-\mu)^{n}}+\sum_{n=0}^{\infty}b_{n}\mu^{n},
\label{eq:36}
\end{equation}
with coefficients $a_{n}$, $b_{n} \in \mathbb{R}$. Expanding the above expression for $\mu\ll 1$ to second order, we must have the following relations for the coefficients $a_{n}$, $b_{n}$ to keep the second condition in \eqref{eq:12}:
\begin{equation}
\begin{split}
b_{0}+\sum_{n=1}^{\infty}a_{n} &= 0,\\
b_{1}+\sum_{n=1}^{\infty}a_{n}n &= 0,\\
b_{2}+\sum_{n=1}^{\infty}a_{n}\frac{n(n+1)}{2} &\neq 0.
\end{split}
\label{eq:37}
\end{equation}
\begin{figure*}
   \centering
   \begin{minipage}[t]{8.5 cm}
\begin{center} 
\includegraphics[trim=20 0 0 0,width=\textwidth]{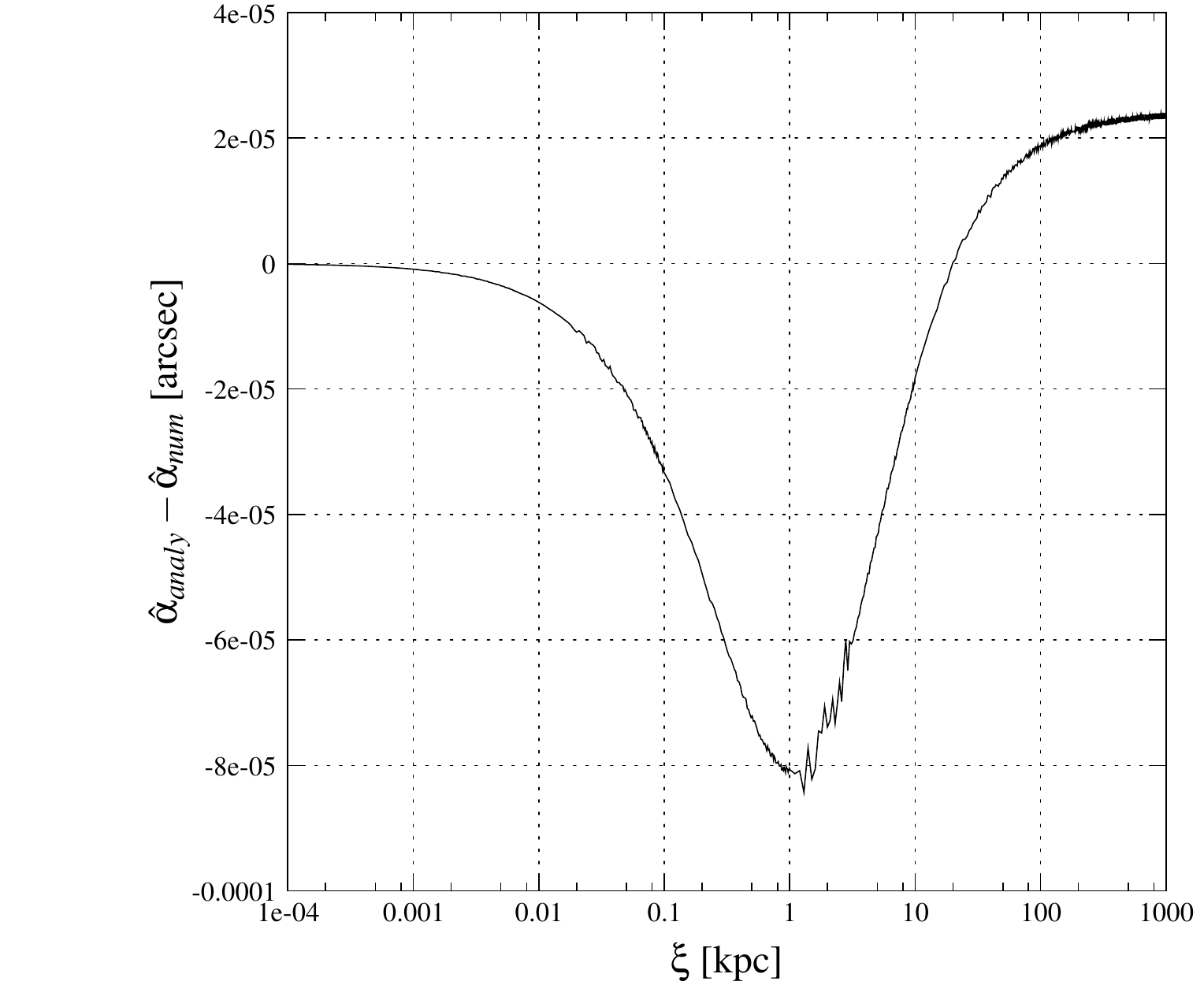}
\end{center}
  \end{minipage}
\qquad
 \begin{minipage}[t]{8.5 cm}
\begin{center}
\includegraphics[trim=50 0 0 0,width=\textwidth]{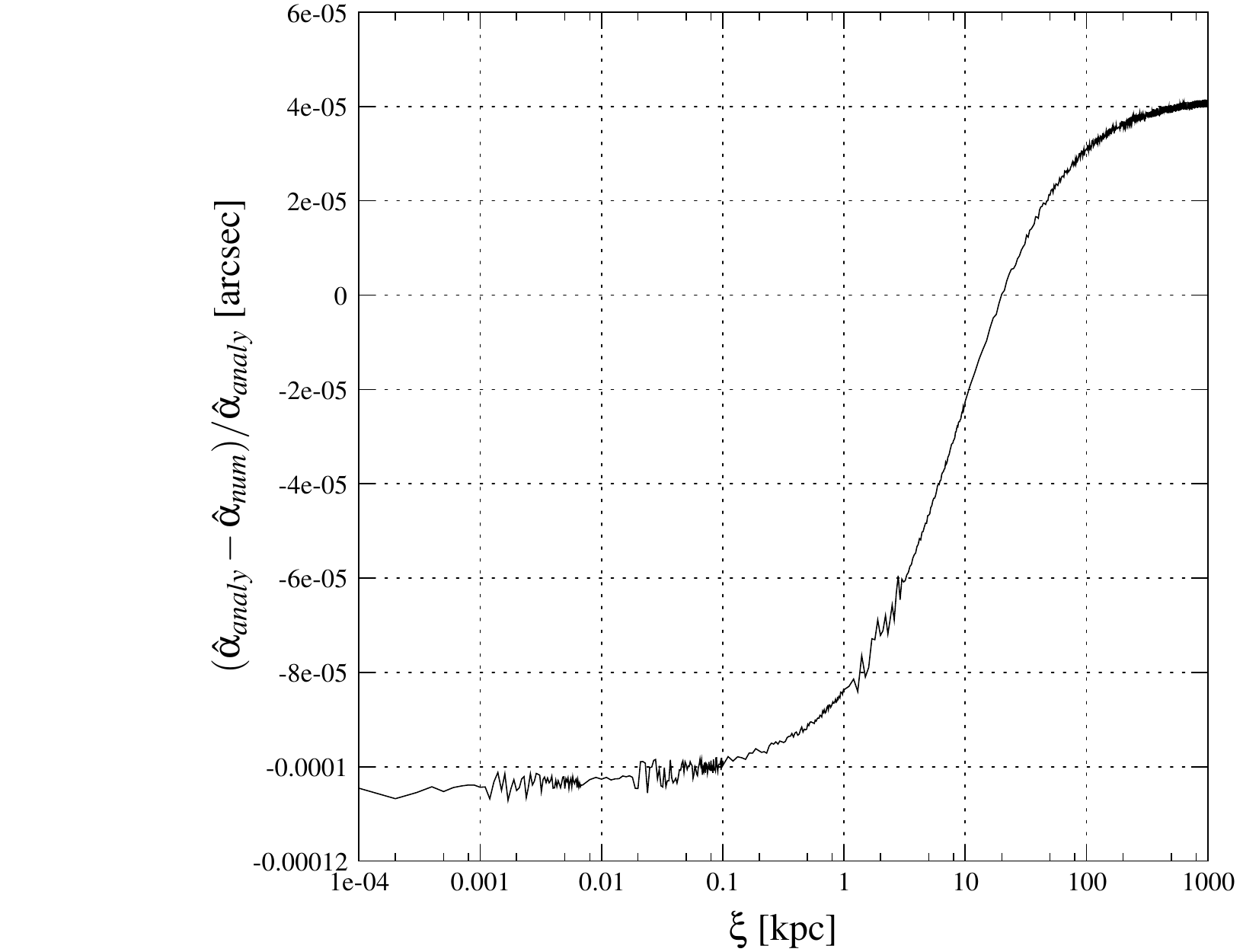}
\end{center}
\end{minipage}
\caption{Absolute (left) and relative (right) difference between the TeVeS deflection angles of the analytic and the numerical Hernquist lens choosing the parameters from Sec. \ref{section33}: Clearly, the deviations are small, $\Delta\hat\alpha\lesssim (10^{-4})^{''}$ and $\Delta\hat\alpha/\hat\alpha_{analy} \lesssim (10^{-4})$. For the numerical calculation, we have assumed $k=0.01$ and $y(\mu)=\mu^{2}/(1-\mu)$.}
\label{comp}
    \end{figure*}
As a simple example, we take the function
\begin{equation}
y(\mu) =\frac{\mu^{2}}{1-\mu}
\label{eq:38}
\end{equation}
and find that the non-zero coefficients are given by
\begin{equation}
a_{1} = 1,\quad b_{0}=-1,\quad b_{1}=-1.
\label{eq:39}
\end{equation}
Setting the coefficients $a_{n}$ and $b_{n}$, we are able to directly control the specific transition behavior from MONDian to Newtonian dynamics. Using the expansion \eqref{eq:36}, we shall study the effects of varying the free function $y(\mu)$ on the deflection angle within numerical analysis.
\subsection{Comparison to the Analytic Model}
\label{comparison}
Taking the simple choice \eqref{eq:38}, we compare the numerical result of the TeVeS Hernquist lens to the analytic solution derived in Sec. \ref{section33}. Fig. \ref{comp} shows the absolute and relative difference between the deflection angles of the analytic and the numerical lens models using the same parameters as in Sec. \ref{section33}, where we have set $k=0.01$ for the numerical calculation. Although we have considered two different free functions $y(\mu)$, the actual differences are fairly small, $\Delta\hat\alpha\lesssim (10^{-4})^{''}$ and $\Delta\hat\alpha/\hat\alpha_{analy} \lesssim (10^{-4})$, and thus negligible with respect to today's observational resolution limit.

Considering the above, it appears that we could be able to determine classes of free functions $y(\mu)$ that nearly produce the same deflection angle. Although we do expect an infinite number of such classes, a closer look will be of advantage (cf. Sec. \ref{choice}). For a systematic approach, we shall make use of the parameterization introduced in Sec. \ref{parameter} to analyze the effects of a varying free function.
\subsection{Varying Parameters}
\label{influence}
Still considering deflection by the Hernquist lens, we now focus on a varying form of the free function $y(\mu)$ and the value of the coupling constant $k$.
For our investigation, all remaining parameters are chosen as in Sec. \ref{section33} unless specified in any other way.

Starting from an arbitrary set ($a_{n},b_{n}$), e.g. the set \eqref{eq:39}, we begin with a variation of the constants $a_{n}$ where we have to adjust $b_{0}$, $b_{1}$ and $b_{2}$ according to \eqref{eq:37}, setting $k$ to a fixed value, e.g. $k=0.01$. Interestingly, numerical analysis has shown that there are no significant changes to the deflection angle for a wide range of parameters, i.e. the relative changes are comparable to those we found in the last section, $\Delta\hat\alpha_{rel} \lesssim 10^{-5}-10^{-4}$. For instance, if we choose
\begin{equation}
a_{18}=1,\quad a_{19}=-2,\quad a_{20}=1,
\label{eq:40}
\end{equation}
which is the expansion of
\begin{equation}
y(\mu) = \frac{\mu^{2}}{(1-\mu)^{20}},
\label{eq:41}
\end{equation}
the relative deviation is of $\mathcal{O}(10^{-5})$. As we have seen, the singularity at $\mu=1$ recovers the Newtonian limit in quasi-static situations, but remarkably, the transition behavior seems almost insensitive to the particular realization of this singularity.

Similarly, we have also examined the effect of a changing coupling constant $k$ taking the coefficients $a_{n}$, $b_{n}$ as constants. Again, the relative differences turned out to be very small, $\Delta\hat\alpha_{rel} \lesssim 10^{-5}-10^{-4}$, varying $k$ within the range of $10^{-4}-10^{-2}$ for different sets ($a_{n},b_{n}$). Obviously, as long as it is small, i.e. $k \lesssim 0.01$, the calculation of the deflection angle does not really depend on the exact value of $k$.

As for the coefficients $b_{n}$ with $n>2$, however, there is a strong influence on the deflection angle, basically allowing to create arbitrary transitions from MOND to Newtonian dynamics. In accordance with the above analysis, it seems that the $b_{n}$ alone can be used to characterize the free function. In general, the exact form of $y(\mu)$ has to by constrained by observational data being independent of the particular law of gravity, which is subject to other work, e.g. \cite{freefunc}.

\section{Non-spherical Lens Models}
Within this section, we will examine the properties of more general lens systems using numerical methods. Introducing our algorithm for the treatment of non-spherical lenses in TeVeS, we will investigate a set of different matter distributions including a toy model of the cluster merger $1$E$0657-558$.
\subsection{Choice of the Free Function}
\label{choice}
Setting $k=0.01$, we shall restrict all further analysis to the following form of $y(\mu)$:
\begin{equation}
y(\mu) = \frac{\mu^{2}}{(1-\mu)^{2}}.
\label{eq:42}
\end{equation}
We will make use of this specific $y(\mu)$ for two reasons: First of all, the choice \eqref{eq:42} is easily inverted, i.e.
\begin{equation}
\mu(y) = \frac{\sqrt{y}}{1+\sqrt{y}},
\label{eq:43}
\end{equation}
and therefore it is possible to express the derivative with respect to $y$ analytically:
\begin{equation}
\frac{\partial\mu}{\partial y} = \frac{1}{2\sqrt{y}(1+\sqrt{y})^{2}}.
\label{eq:44}
\end{equation}
As will become clear in Sec. \ref{sectionsolver}, both $\mu(y)$ and $\partial\mu/\partial y$ are part of Eq. \eqref{eq:15}. Since $\partial\mu/\partial y \rightarrow \infty$ for $y\rightarrow 0$, a possible solver of \eqref{eq:15} might be extremely sensitive to the corresponding run of $\partial\mu/\partial y$ in that regime. By choosing the analytic expressions \eqref{eq:43} and \eqref{eq:44}, respectively, we are able to avoid numerical inversion and differentiation of the free function, which may prevent a destabilizing influence on the algorithm.

Secondly, our choice allows us to use the analytic Hernquist lens for comparison in order to test the accuracy of a non-spherical solver for this specific density profile. According to Sec. \ref{spherical}, Eq. \eqref{eq:42} is close to the choice \eqref{eq:30} and produces nearly the same deflection angle, thus justifying such a comparison.
\subsection{Calculating the Scalar Potential}
\label{sectionsolver}
Since $\mu = \mu(y)$ and $y=kl^{2}|\vec\nabla\phi|^{2}$ for quasi-static systems, an expansion of the l.h.s. of Eq. \eqref{eq:15} yields
\begin{equation}
2\frac{\partial\mu}{\partial y}kl^{2}\left((\partial_{i}\phi)(\partial_{j}\phi)(\partial_{i}\partial_{j}\phi)\right)+\mu\Delta\phi = kG\rho.
\label{eq:45}
\end{equation}
Defining an effective matter density $\bar\rho$ such that
\begin{equation}
\Delta\phi = \bar\rho,
\label{eq:46}
\end{equation}
where
\begin{equation}
\bar\rho = \frac{kG}{\mu}\rho-2\frac{kl^{2}}{\mu}\frac{\partial\mu}{\partial y}\left((\partial_{i}\phi)(\partial_{j}\phi)(\partial_{i}\partial_{j}\phi)\right),
\label{eq:47}
\end{equation}
we may choose an appropriate first guess of $\phi$ and calculate an initial density $\bar\rho^{(0)}$ by using Eq. \eqref{eq:47}. Solving Poisson's equation by means of Fourier methods, i.e. Eq. \eqref{eq:46} with the r.h.s. being fixed ($\bar\rho = \bar\rho^{(0)}$), we find a new field $\phi^{(1)}$, which can be used to obtain $\bar\rho^{(1)}$ and so forth.

Without any further modification, this approach fails to converge in most cases, with the $\phi^{(n)}$ oscillating rapidly. Including a relaxation into the iteration, however, it is possible to enforce convergence for a variety of problems, and thus our final iterative scheme reads as ($\bar\rho^{(0)}$ is calculated from an initial guess $\phi_{0}$)
\begin{equation}
\begin{split}
\Delta\tilde{\phi}^{(n)} &= \bar\rho^{(n)},\\
\phi^{(n+1)} = \omega\tilde\phi^{(n)}&+(1-\omega)\phi^{(n)},
\end{split}
\label{eq:48}
\end{equation}
where we have introduced the relaxation parameter $\omega\in\mathbb{R}$, an additional iteration field $\tilde\phi^{(n)}$ and
\begin{equation}
\begin{split}
\bar\rho^{(n)} &= \frac{kG}{\mu^{(n)}}\rho-2\left(\frac{\partial\mu}{\partial y}\right)^{(n)}\frac{kl^{2}}{\mu^{(n)}}\left((\partial_{i}\phi^{(n)})(\partial_{j}\phi^{(n)})(\partial_{i}\partial_{j}\phi^{(n)})\right),\\
\mu^{(n)} &= \mu(y^{(n)}), \quad \left(\frac{\partial\mu}{\partial y}\right)^{(n)}=\frac{\partial\mu}{\partial y}(y^{(n)}), \quad y^{(n)} = kl^{2}|\vec\nabla\phi^{(n)}|.
\end{split}
\label{eq:49}
\end{equation}
For suitable values of $\omega$, our method turns out to work very well for a wide range of density profiles (cf. Sec. \ref{problems}). { However, our investigation has shown that the relaxation's success is very sensitive to the particular choice of $\omega$, i.e. $\omega$ has to be chosen from a very narrow range, $\omega=0.75\pm0.1$. Although convergence is achieved within a wider range of $\omega$, its behavior quickly deteriorates. Fortunately, this value seems} to be almost independent of the particular density profile, and therefore it will not be necessary to adjust $\omega$ once it has been determined for a certain density.

\subsubsection{Point Lens Approximation}
\label{pointlens}
As the scalar field's gradient decreases much more slowly compared to the Newtonian one far away from the lens, one would actually be obliged to move to very large volumes in order to neglect contributions from outside the box and obtain correct results for the deflection angle. Thus, assuming a fixed grid size, this would excessively degrade the resolution of the corresponding two-dimensional lensing maps. In the following, we shall discuss an approximation allowing us to avoid this problem:

Considering a finite grid with $N+1$ points per dimension ($N$ is chosen as an even number), we may rewrite the scalar part of the deflection angle as the sum of contributions coming from both inside and outside the grid's volume:
\begin{equation}
\vec{\hat\alpha}_{s}=2\int_{-\frac{N}{2}\Delta x}^{\frac{N}{2}\Delta x}\vec\nabla_{\bot}\phi^{(in)}dz+4\int_{\frac{N}{2}\Delta x}^{\infty}\vec\nabla_{\bot}\phi^{(out)}dz,
\label{eq:50}
\end{equation}
with the quantity $\Delta x$ denoting the distance between neighboring grid points. Assuming that the scalar field at the boundaries is approximately given by that of a point lens, i.e.
\begin{equation}
\phi^{(out)} \approx \sqrt{GMa_{0}}\log(r),
\label{eq:51}
\end{equation}
we obtain the following expression ($M$ denotes the total mass inside the volume):
\begin{equation}
\vec{\hat\alpha}_{s}=2\int_{-\frac{N}{2}\Delta x}^{\frac{N}{2}\Delta x}\vec\nabla_{\bot}\phi^{(in)}dz+4\vec A,
\label{eq:52}
\end{equation}
where
\begin{equation}
\vec A = \frac{\sqrt{GMa_{0}}}{q}\left\lbrack\frac{\pi}{2}-\arctan\left(\frac{N\Delta x}{2q}\right)\right\rbrack
\begin{pmatrix}
x\\
y
\end{pmatrix},
\label{eq:53}
\end{equation}
and $q^{2} = x^{2} + y^{2}$.
If applicable, we need to perform the integration only over our finite grid since all contributions from outside the box can be expressed analytically. Concerning our iterative solver, we may additionally assume the boundary conditions of the fields $\phi^{(n)}$ to be of spherical symmetry, and it turns out to be sufficient to use \eqref{eq:51} as an initial guess for $\phi$. { The highest resolution compatible with the limitations of our computer hardware is set by $N=384$ which is used for all numerical calculations}. Before turning to non-spherical lens systems, however, we shall examine our method's accuracy.
\subsubsection{Accuracy}
\label{accuracy}
Comparing the numerically obtained deflection angle of a Hernquist lens to the analytic result \eqref{eq:32}, we will determine the accuracy of our tool assuming the parameters from Sec. \ref{section33}, which corresponds to a galaxy-sized mass distribution. As previously mentioned, such a comparison is justified according to our analysis in Sec. \ref{parameter}. Concerning the numerical setup, we choose a grid volume of $V = (50$kpc$)^{3}$ (the lens is placed in the grid's center), and in order to obtain a sufficiently large value of $D=D_{ds}D_{d}/D_{s}$, we set the redshifts of source and lens to $z_{source}=3$ and $z_{lens}=0.63$, 
\begin{figure*}[t]
\begin{minipage}[t]{9cm}
\begin{center} 
\includegraphics[width=10cm]{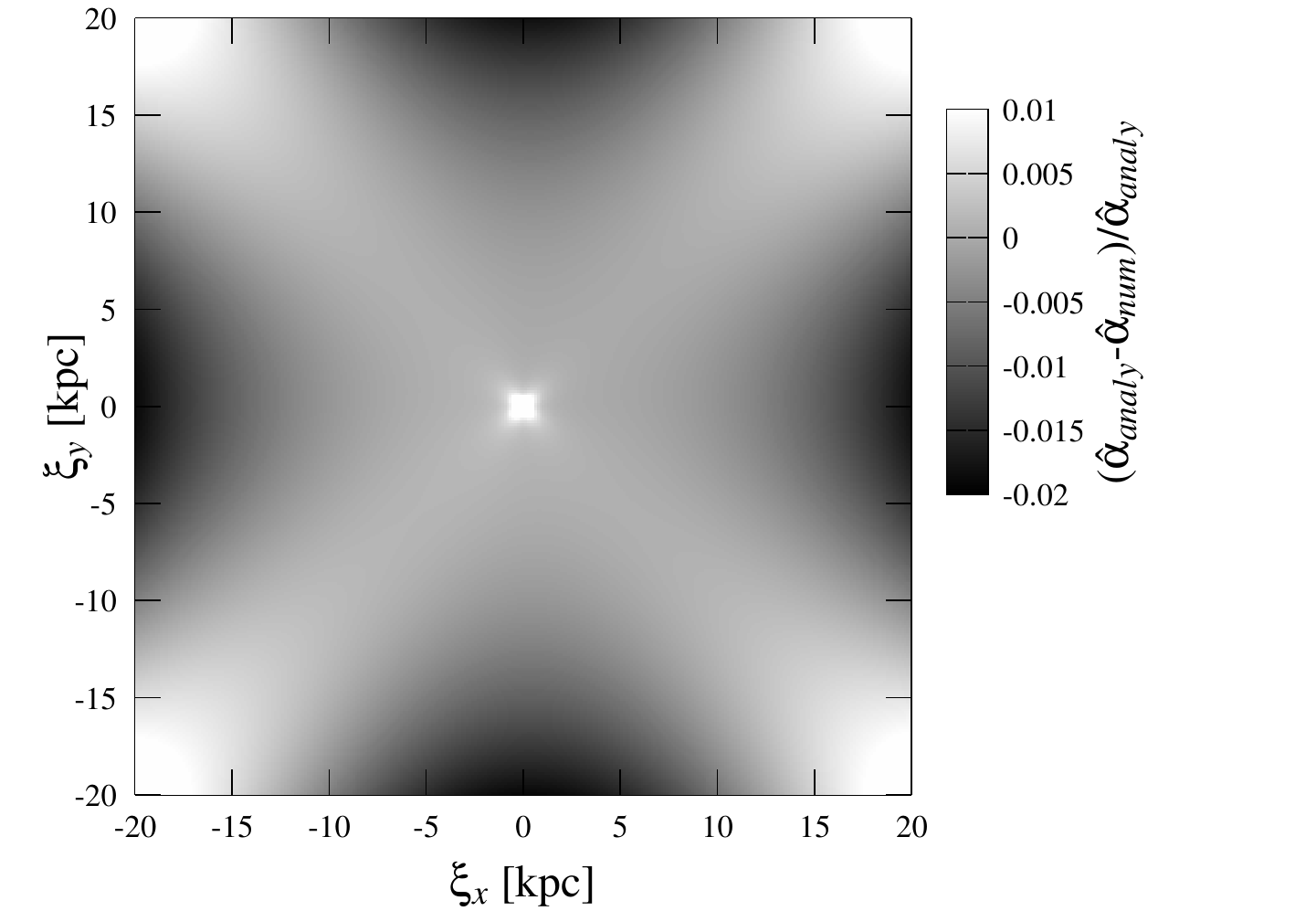}
\end{center}
  \end{minipage}
\hfill
 \begin{minipage}[t]{9cm}
\begin{center}
\includegraphics[width=10cm]{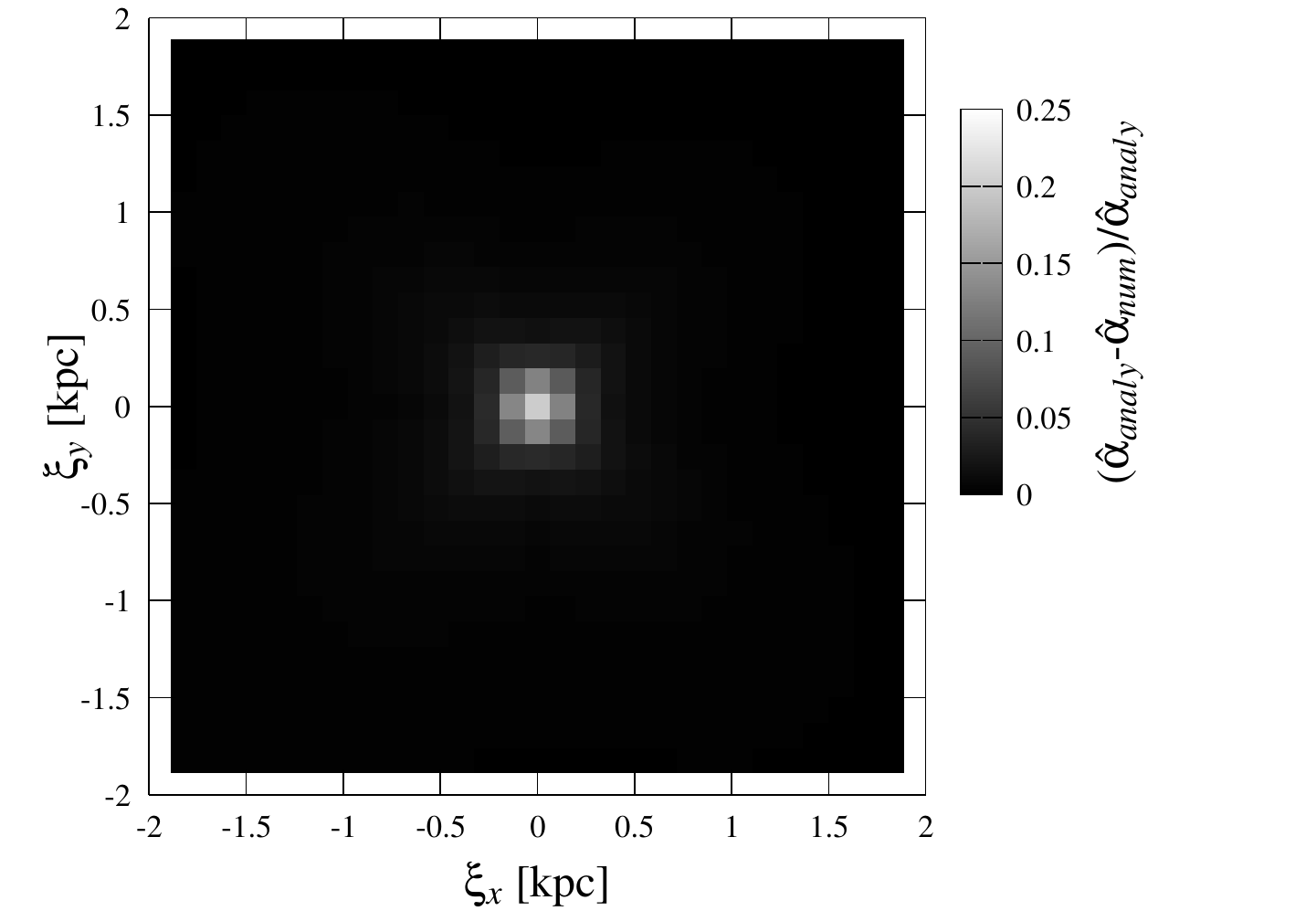}
\end{center}
\end{minipage}
\caption{Accuracy test of our numerical method: Assuming a Hernquist lens with the parameters from Sec. \ref{section33}, we compare the numerical result for the deflection angle $\hat\alpha(\vec\xi)$ to the analytic one. The left panel shows the relative difference $\Delta\hat\alpha_{rel}=(\hat\alpha_{analy}-\hat{\alpha}_{num})/\hat\alpha_{analy}$. Please note that the quantity $\Delta\hat\alpha_{rel}$ is limited by $-0.02\leq\Delta\hat\alpha_{rel}\leq0.01$ for reasons of presentation, values outside this range are truncated. The right panel illustrates the relative deviation for the central part where $\Delta\hat\alpha_{rel}$ reaches a maximum of approximately $20\%$.}
\label{acc}
\end{figure*}
respectively. If not explicitly noted, we shall keep this choice of redshifts throughout the following sections. At $\xi\approx25$kpc, the relative deviation between the analytic Hernquist and its corresponding point lens is approximately $6\%$. Although this difference is quite large, we assume the validity of Eq. \eqref{eq:51} in order to balance accuracy and resolution ($N=384$). Thus, we are able to resolve structures up to a minimum extent of $\Delta x\approx130pc$.

The left panel of Fig. \ref{acc} shows the relative difference $\Delta\hat\alpha_{rel}=(\hat\alpha_{analy}-\hat{\alpha}_{num})/\hat\alpha_{analy}$ between the numerical and the analytic deflection angle of the Hernquist lens. For reasons of presentation, we have limited the range of $\Delta\hat\alpha_{rel}$ to $-0.02\leq\Delta\hat\alpha_{rel}\leq0.01$. Ignoring the very center of the map (right panel), we find the relative deviations in the interior
are of $\mathcal{O}(10^{-3})$. Moving outwards, i.e. to larger $\xi=|\vec\xi|$, these deviations increase and reach values up to $5-6\%$ at the grid's boundaries ($\xi\gtrsim25$kpc). However, as long as $\xi\lesssim15$kpc, we still have $\Delta\hat\alpha_{rel}\lesssim1\%$, again neglecting the central part. The large differences close to the boundaries are likely to be a mixture of artifacts caused by the Fourier transform of actual non-periodic fields and contributions due to \eqref{eq:53} which become more significant with increasing $\xi$.
Having a look at the right panel of Fig. \ref{acc}, we see that $\Delta\hat\alpha_{rel}$ strongly increases in the central region reaching a maximum value of roughly $20\%$. The reasons for these large deviations are probably related to both the limited resolution of our grid and the small values of $\hat\alpha$ in the center { (see also Sec. \ref{problems})}. According to Sec. \ref{section33}, the TeVeS deflection angle of the analytic Hernquist lens decreases to zero for $\xi\rightarrow\ 0$. Since this transition happens on a rather small scale, our numerical model cannot fully recover the deflection angle in the central region. Moreover, we have to consider that the matter density \eqref{eq:31} becomes infinite at $\xi=0$, which, of course, cannot be accomplished in a numerical calculation.  Due to the grid, this singularity is smoothed out, causing an effective loss of mass in our numerical model. This loss has an overall influence on the deflection angle and may significantly contribute to the errors we have discussed above. Investigating non-spherical systems, however, we shall only consider lenses which follow smooth density distributions.

For numerical simulations that similarly allow using the point lens approximation, we may assume an accuracy equal to that of the Hernquist lens. As we are mainly interested in the strong lensing regime, we shall restrict ourselves to the grid's interior where the relative deviations are of $\mathcal{O}(10^{-3})$. Due to finite resolution, however, we expect the accuracy to degrade to some extent in regions where the deflection angle $\hat\alpha$ approaches values close to zero when moving to more generic lens systems. Although smooth density profiles will probably not produce deviations as large as we have found around $\xi= 0$ for the Hernquist lens, we cannot make any specific statements on the quality of our simulations in such areas. Still, this should affect but a fraction of the overall result, thus being acceptable for the following analysis.
\begin{figure*}[t]
\begin{minipage}[t]{9cm}
\begin{center} 
\includegraphics[width=9cm]{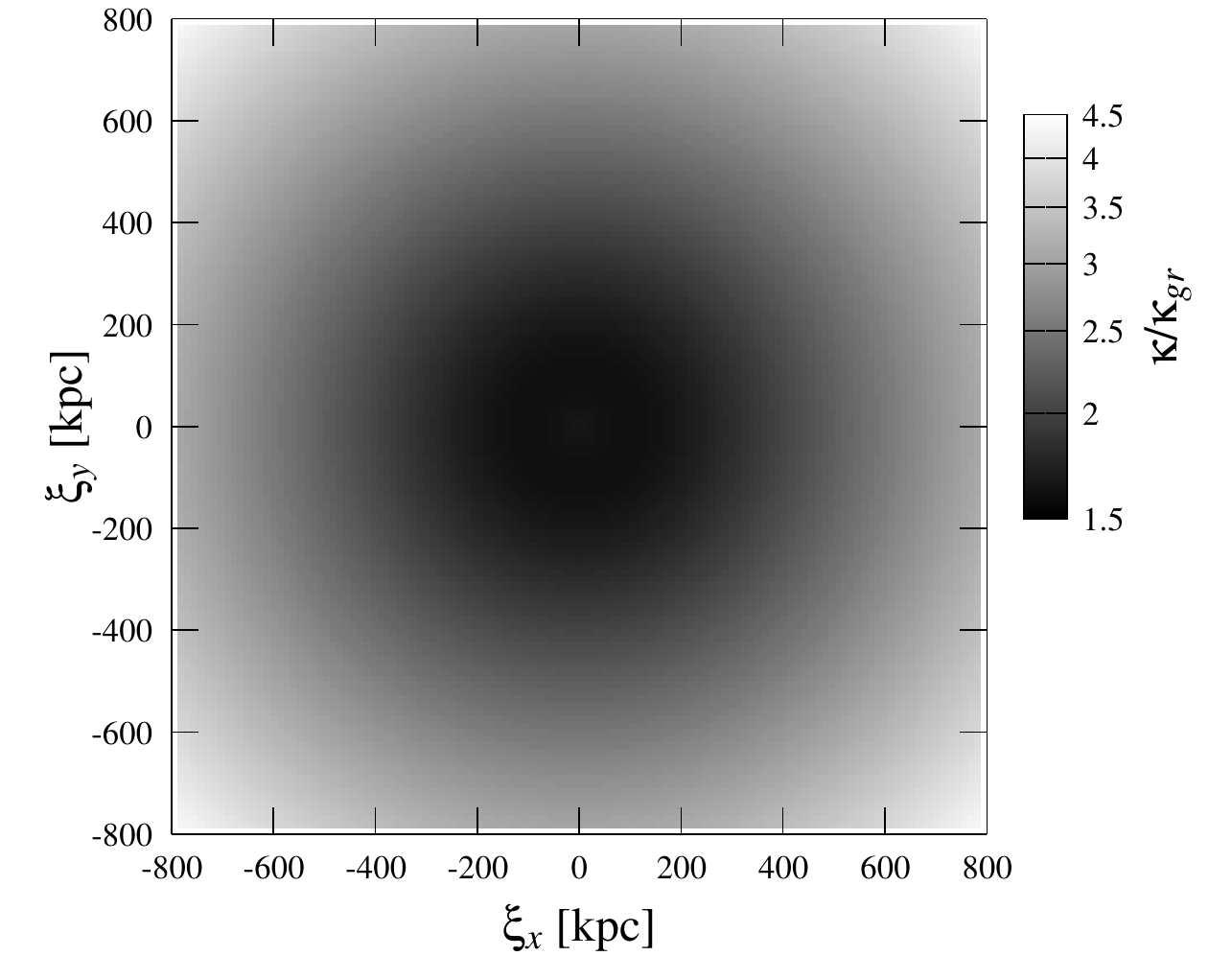}
\end{center}
  \end{minipage}
\begin{minipage}[t]{1cm}
\begin{center} 
\end{center}
  \end{minipage}
 \begin{minipage}[t]{10.2cm}
\begin{center}
\includegraphics[width=10.2cm]{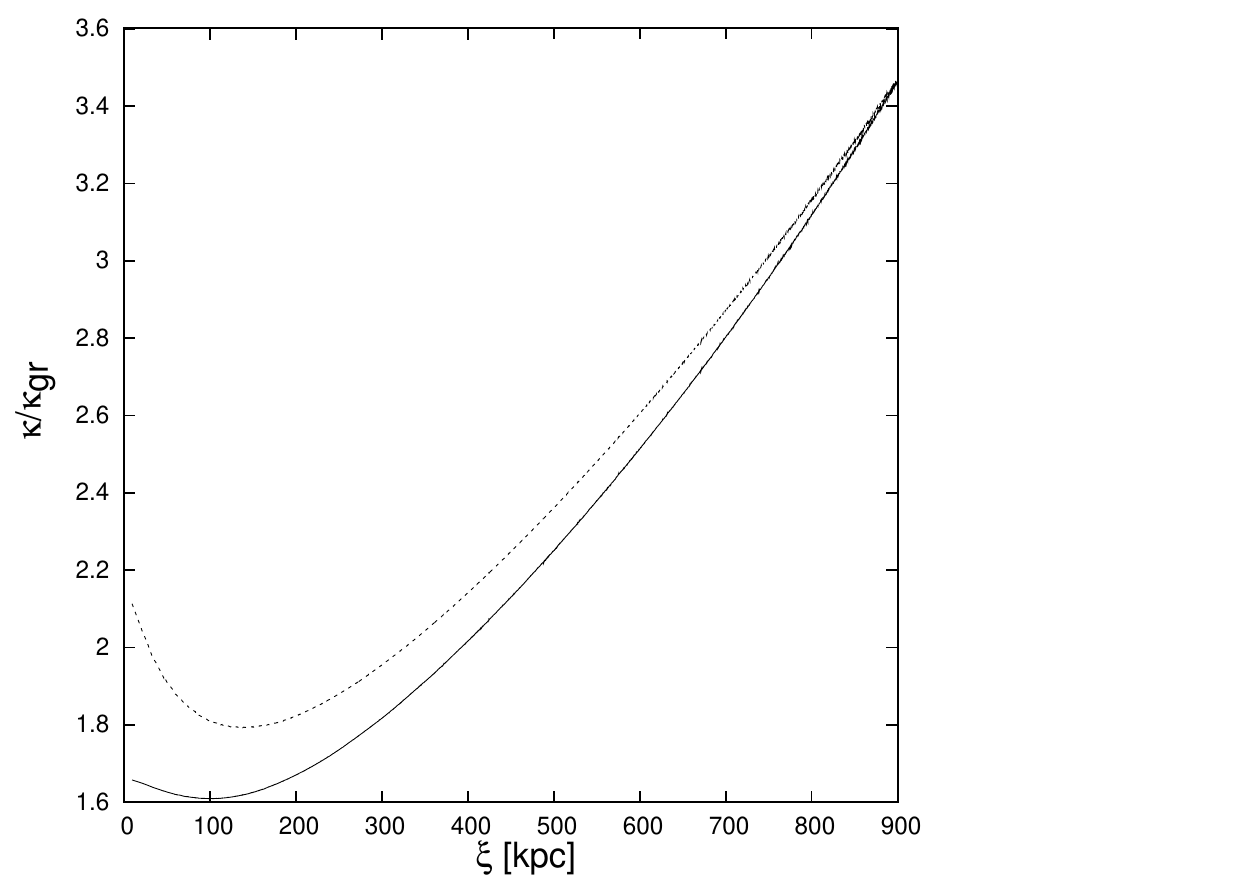}
\end{center}
\end{minipage}
\caption[Numerically calculated TeVeS convergence map for the King-like profile expressed in terms of the GR convergence $\kappa_{gr}$]{Left: Numerically calculated TeVeS convergence map for the King-like profile \eqref{eq:54} expressed in terms of the GR convergence $\kappa_{gr}$ assuming $z_{0}=50$kpc. Right: Since \eqref{eq:54} corresponds to an axisymmetric configuration, the effective TeVeS convergence $\kappa$ and the ratio $\kappa/\kappa_{gr}$ depend on the radial coordinate $\xi$ only. As the calculated convergence maps relatively deviate from circular symmetry by $\mathcal{O}(10^{-3})$, which is due to our Fourier method, the presented results are averaged over all directions. Choosing $z_{0}=50$kpc (solid) and $z_{0}=400$kpc (dashed), we see that $\kappa$ is significantly amplified in the central region when moving to higher values of $z_{0}$.}
\label{thin1}
\end{figure*}

\subsubsection{Problems}
\label{problems}
As for the solver of the scalar field, { we have encountered the following problems:} Considering more complicated density distributions, we have found the relaxed iteration to be less efficient, i.e. the iteration generally takes more time to converge. As it turns out, this cannot be compensated by changing the relaxation parameter $\omega$, which would actually lead to even worse convergence properties or a complete failure of the method. Still, the additional amount of time that has to be employed is acceptable in most cases { and corresponds to a factor of $2-3$}.

Furthermore, independently of the particular value of $\omega$, we encounter the relaxation to generally fail for certain choices of $\rho$. Since it can mostly be resolved by slightly modifying the original density profile, this second problem is probably of purely numerical origin. However, we point out that it might also hint on an exceptional behavior of the scalar field $\phi$ that is not accessible to our solver. { Considering the Hernquist profile, for instance, the difficulties found in the central part may reflect its intrinsic instability w.r.t. TeVeS/MOND, rather than a negative feature of our code \citep{convergence}.}

\subsection{Thin Lens Approximation}
\label{thinlens}
As our first task, we want to investigate the validity of the thin lens approximation in TeVeS. According to former work considering lensing in classical MOND \citep{thinfail}, we expect a break-down of the approximation due to the non-linear coupling of the scalar field to the three-dimensional matter density. In the following, however, we are rather interested in quantifying this break-down by exploring the lensing properties of a mass distribution being contracted or stretched along the line of sight, i.e. the $z$-direction, making use of our new numerical tool. For this purpose, let us consider a three-dimensional density distribution $\rho$ following a King profile { \citep{king}} which is given by
\begin{equation}
\rho(r) = \rho_{0}\left(1+\left(\frac{r}{r_{c}}\right)^{2}\right)^{-\frac{3}{2}},
\label{eq:54}
\end{equation}
where $r_{c}$ is the core radius and $\rho_{0}$ the matter density at $r=0$. Eq. \eqref{eq:54} is an empirical law that fairly describes the distribution of both galaxies and gas inside a galaxy cluster. However, in order to analyze TeVeS effects which are only due to the lens's extent along the line of sight, we have to parameterize its ``thickness" and additionally ensure a constant projected mass density. Thus, we introduce a slightly modified profile:
\begin{equation}
\rho(q,z) = \rho_{0}\mathcal{Q}(q)\mathcal{Z}(z),
\label{eq:55}
\end{equation}
where
\begin{equation}
\mathcal{Q}(q) = \left(1+\left(\frac{q}{q_{0}}\right)^{2}\right)^{-1}
\label{eq:56}
\end{equation}
and
\begin{equation}
\mathcal{Z}(z) = \left(1+\left(\frac{z}{z_{0}}\right)^{2}\right)^{-\frac{1}{2}},
\label{eq:57}
\end{equation}
with $q_{0}$, $z_{0}>0$ being the corresponding core lengths and $q^{2}=x^{2}+y^{2}$. Since the expressions \eqref{eq:56} and \eqref{eq:57} are obtained by integrating \eqref{eq:54} over one and two dimensions, respectively, our new choice \eqref{eq:55} is actually kept close to the original King profile. Varying the parameter $z_{0}$, we are now able to directly control the lens's extent in the $z$-direction.

\begin{figure}
   \centering
   \includegraphics[trim=82 0 0 0,width=\linewidth]{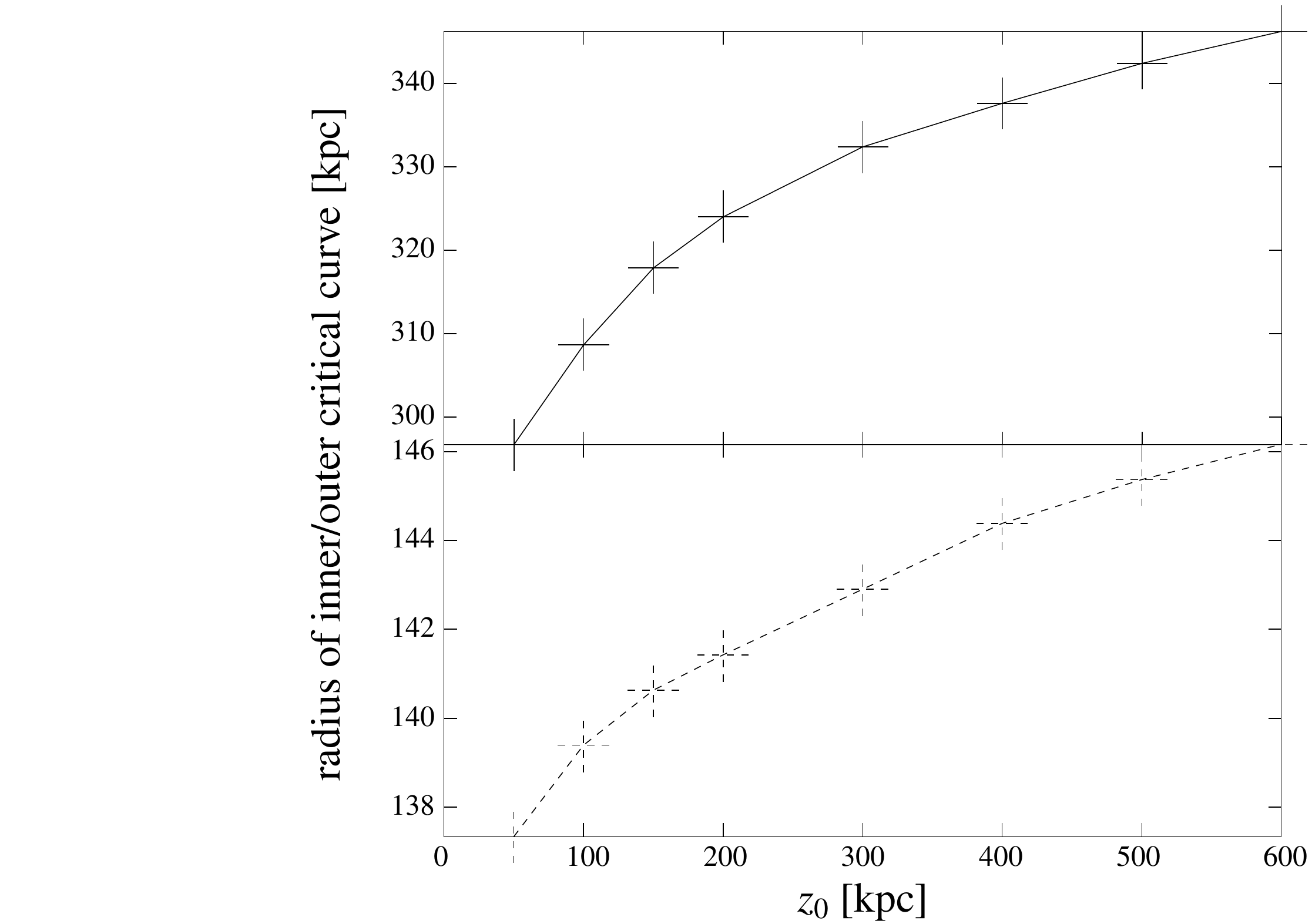}
\caption[Radii of the inner and outer critical curves for different choices of $z_{0}$]{Radii of the inner (dashed) and outer (solid) critical curve for different choices of $z_{0}$: Increasing the lens's extent along the line of sight, the radius of the inner (outer) critical curve is driven outwards showing relative changes of up to roughly $6\%$ $(16\%)$. The critical lines are calculated by interpolation between the grid points.}
         \label{thin2}
   \end{figure}

\begin{figure*}[!t]
\begin{minipage}[t]{8.5cm}
\begin{center} 
\includegraphics[width=8.5cm]{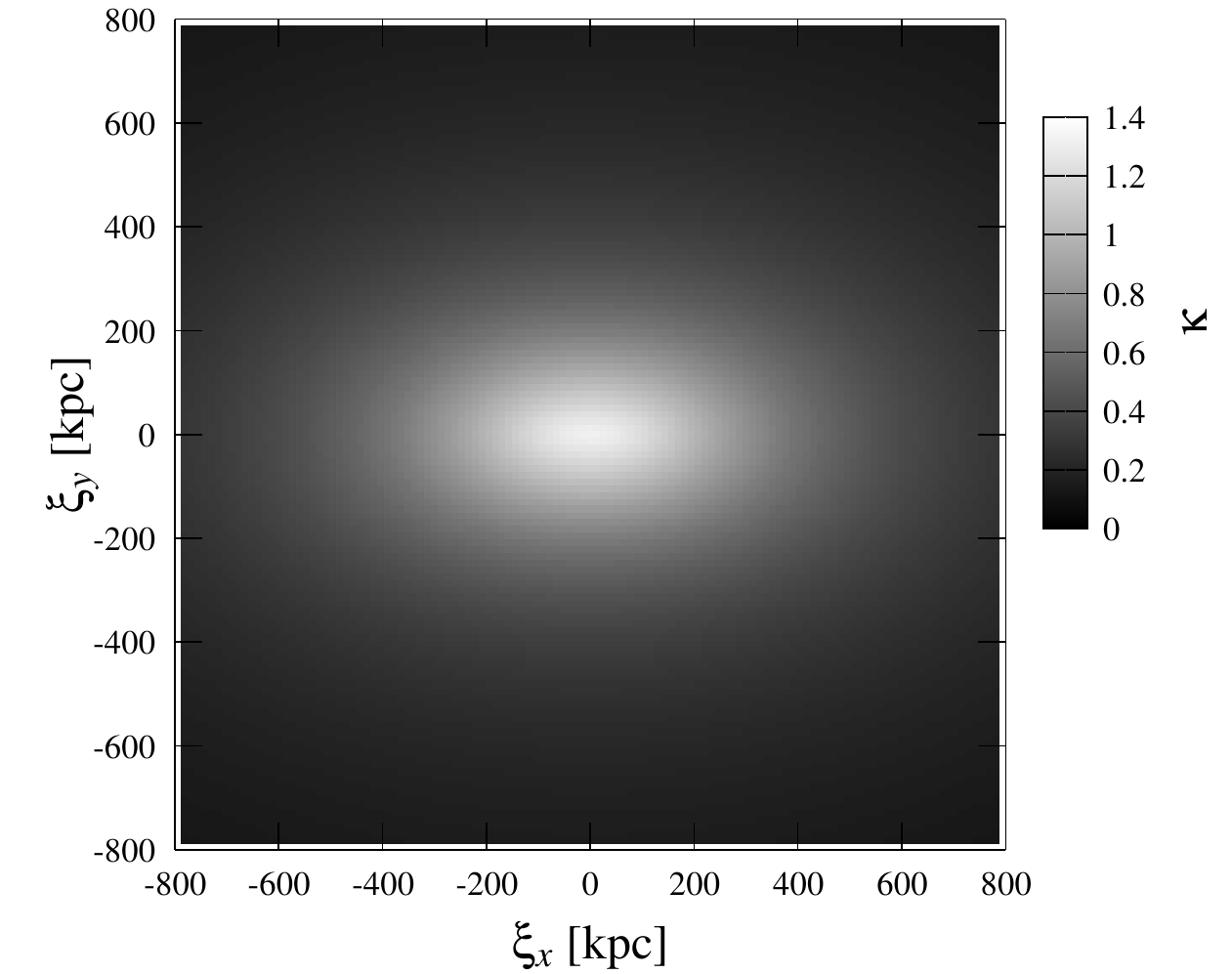}
\end{center}
  \end{minipage}
\hfill
 \begin{minipage}[t]{8.5cm}
\begin{center}
\includegraphics[trim= 10 0 0 0,width=8.5cm]{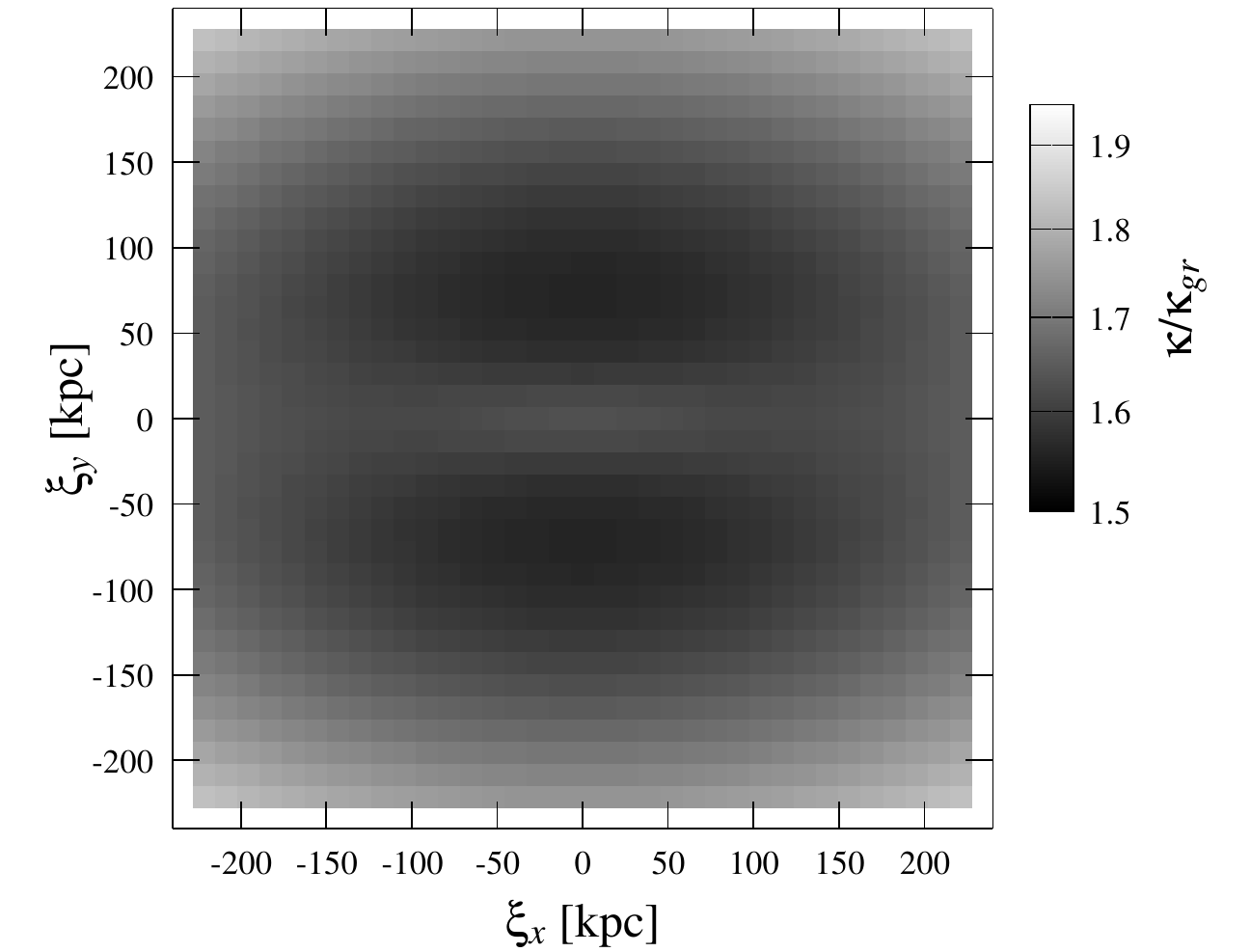}
\end{center}
\end{minipage}
\caption{Numerically calculated TeVeS convergence $\kappa$ (left) and the corresponding ratio $\kappa/\kappa_{gr}$ (right) for an elliptical profile of type \eqref{eq:58} with $x_{0}=350$kpc and $y_{0}=z_{0}=200$kpc: In the central part, the TeVeS convergence $\kappa$ is unevenly amplified, thus breaking the elliptical symmetry. Compared to the dominant GR contributions, however, this effect is almost negligible.}
\label{ellipt1}
\end{figure*}
Concerning the numerical setup, we take the point of origin, i.e. $x=y=z=0$, to coincide with the grid's center and assume the density profile \eqref{eq:55} to be smoothly cut at a radius of $R=1.5$Mpc, which is necessary as, otherwise, our King-like profile would contain an infinite amount of mass. In accordance with the point lens approximation, we set the grid's volume to $V=(5$Mpc$)^{3}$, thus having a spatial resolution of $\Delta x\approx13$kpc $(N=384)$. Furthermore, we take $q_{0}=200$kpc and choose $\rho_{0}$ such that the lens's total mass is given by $M=10^{15}M_{\odot}$, which now corresponds to a cluster-sized mass distribution. Regarding the remaining parameters, we keep the settings introduced in Sec. \ref{section33}, the redshifts of source and lens being fixed to $z_{source}=3$ and $z_{lens}=0.63$, respectively. Henceforth, if not explicitly noted, all presented numerical results are based upon these settings, including the spherical cut-off of the particular density profile at $R=1.5$Mpc. 
As we have to meet condition \eqref{eq:51} in order to apply the point lens approximation, we are obliged to satisfy $z_{0}\lesssim600$kpc in our numerical simulations.

The right panel of Fig. \ref{thin1} shows the effective TeVeS convergence of our King-like profile expressed in terms of the corresponding GR convergence $\kappa_{gr}$ for $z_{0}=50$kpc and $z_{0}=400$kpc, respectively. Note that the GR maps are independent of the particular choice of $z_{0}$. Increasing the value of $z_{0}$, we observe a significant amplification of the TeVeS convergence around the center while there is basically no change in the outer region. As expected, the variation of $z_{0}$ has no effect on the symmetry properties of the convergence map. Concerning the TeVeS shear map, we find a similar behavior: While there is a strong increase of $\gamma$ in the very center, we find only small changes in the outer parts. Interestingly, the TeVeS shear is not exactly circularly symmetric in that region any longer, with the actual form depending on the particular extent of the lens. Rather than being intrinsic to TeVeS, however, this is probably due to Fourier artifacts caused by the scalar field solver or the point lens approximation, an influence of the latter being actually expected as the choice of $z_{0}$ has an impact on Eq. \eqref{eq:51}.

Let us continue our analysis considering the effects on the critical lines due to the changes of $\kappa$ and $\gamma$: Since \eqref{eq:55} is axially symmetric, the corresponding lines turn into circles. In Fig. \ref{thin2}, the radii of both the inner and outer critical curve are presented for different values of the parameter $z_{0}$. Obviously, these radii are increased when stretching the lens along the line of sight, showing relative deviations of up to roughly $6\%$ and $16\%$ for the radial and tangential critical radius, respectively. Note that the critical lines are calculated by interpolation between the grid points, thus allowing to determine positions which are below the grid's resolution. Although our investigation is limited to a small range of $z_{0}$, we find appreciable differences between the lensing maps which are assumed to considerably grow when stretching the lens further.

Summarizing the above, we may conclude that the lens's extent along the line of sight significantly affects the strong lensing properties. Therefore, the mass distribution along the $z$-axis can be regarded as an additional degree of freedom in TeVeS.
\subsection{Elliptical Lenses}
\begin{figure*}[t]
\begin{minipage}[t]{9cm}
\begin{center}
\includegraphics[width=10cm]{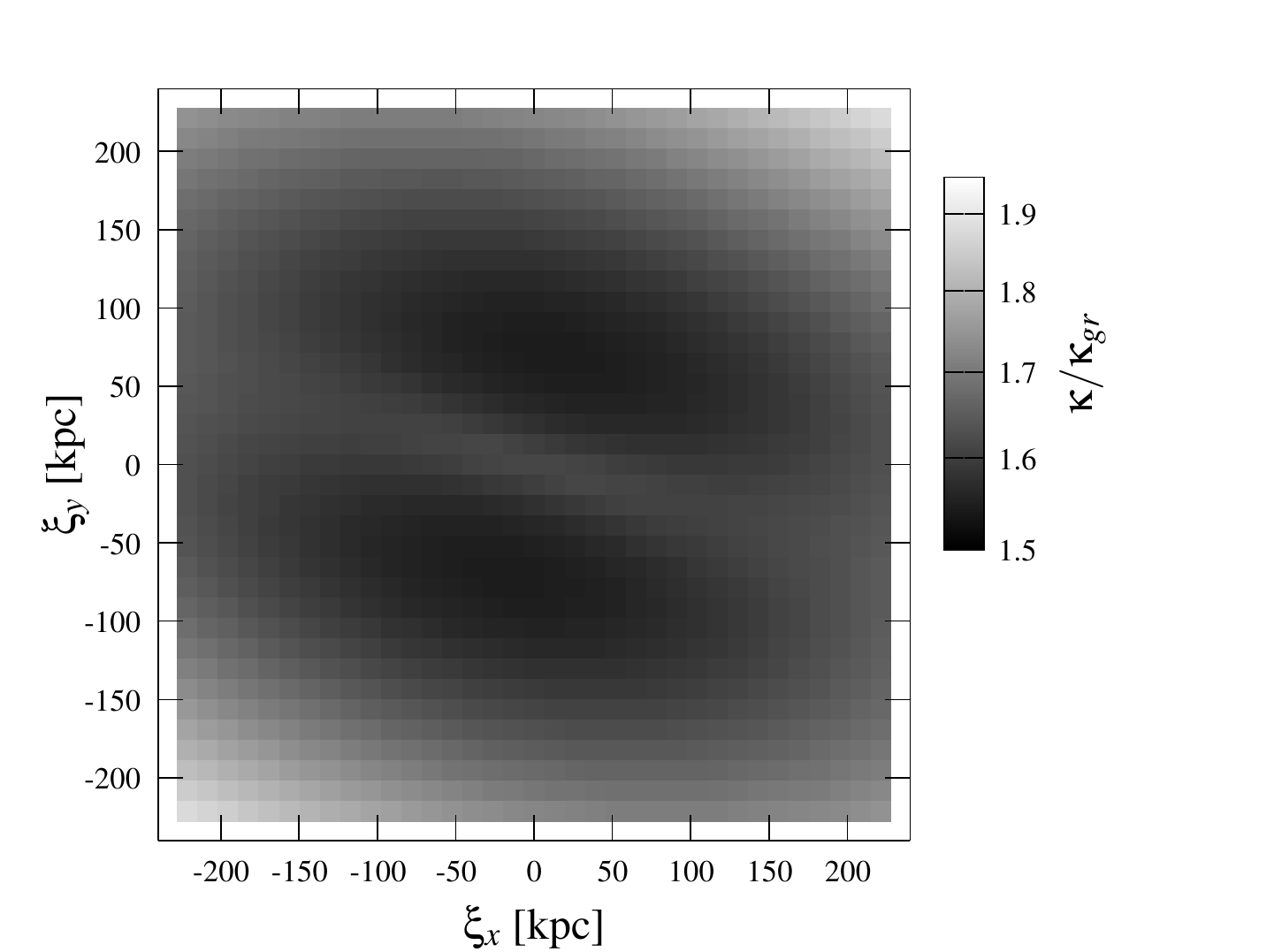}
\end{center}
\end{minipage}
\hfill
\begin{minipage}[t]{9cm}
\begin{center} 
\includegraphics[trim= 2cm 3cm 0 1cm,width=10cm]{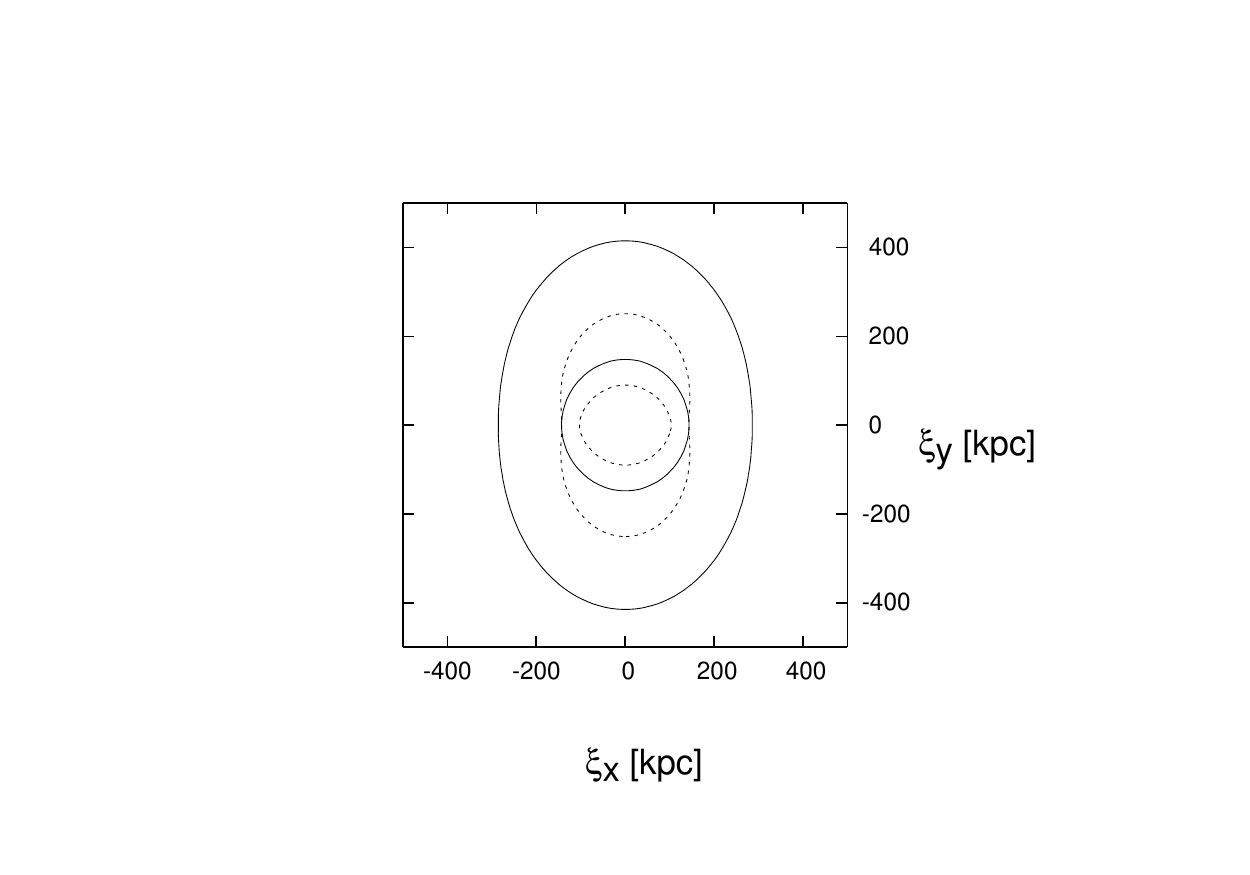}
\end{center}
\end{minipage}
\caption{Left panel: Numerically calculated TeVeS convergence ratio $\kappa/\kappa_{gr}$ for the rotated profile \eqref{eq:58} with $x_{0}=350$kpc: In the central part, the TeVeS convergence $\kappa$ is unevenly amplified, similar to the unrotated case shown in the bottom panel of Fig. \ref{ellipt1}. Thus, it seems unlikely that the observed effect is a numerical artifact caused by our method.
Right panel: Critical curves for both TeVeS (solid) and GR (dashed) assuming an elliptical profile of type \eqref{eq:58} with $x_{0}=150$kpc and $y_{0}=z_{0}=200$kpc.
}
         \label{rotate}
   \end{figure*}

In this section, we shall consider lens systems whose projected mass density follows an elliptic profile. Therefore, introducing the scale lengths $x_{0},y_{0},z_{0}>0$, let us consider a matter density distribution of the form
\begin{equation}
\rho(r) = \rho_{0}\left(1+\left(\frac{x}{x_{0}}\right)^{2}+\left(\frac{y}{y_{0}}\right)^{2}+\left(\frac{z}{z_{0}}\right)^{2}\right)^{-\frac{3}{2}}.
\label{eq:58}
\end{equation}
Keeping $y_{0}$ and $z_{0}$ fixed, $y_{0}=z_{0}=200$kpc, we investigate the lensing properties for different choices of the parameter $x_{0}$, again setting the total mass to $M=10^{15}M_{\odot}$. Fig. \ref{ellipt1} illustrates both the TeVeS convergence $\kappa$ and the corresponding ratio $\kappa/\kappa_{gr}$, with $x_{0}$ set to a value of $350$kpc. Although the symmetry properties of the GR convergence map are virtually sustained in TeVeS, we can observe an interesting feature located in the central part: Compared to its neighborhood, there is a slightly increased amplification close to the semi-major axis, breaking the elliptical symmetry. If this effect was larger, it could actually account for loosing track of the baryonic matter distribution, thus yielding a qualitatively different looking TeVeS convergence $\kappa$. For $x_{0}=350$kpc, we additionally present a simulation where the density profile \eqref{eq:58} has been rotated around the $z$-, $y$- and $x$-axis by $10^{\circ}$, $20^{\circ}$ and $30^{\circ}$, respectively. Clearly, the ratio $\kappa/\kappa_{gr}$ illustrated in Fig. \ref{rotate} (left panel) shows essentially the same inner structure as in Fig. \ref{ellipt1}. Therefore, it seems unlikely that the observed effect is a numerical artifact caused by our method.

In the right panel of Fig. \ref{rotate}, we compare the critical lines in TeVeS to those obtained in GR assuming $x_{0}=150$. The found symmetry-breaking effect does not appear to have any significant influence on the critical curves which therefore do not show any unfamiliar shapes compared to elliptical GR lenses. As the TeVeS convergence $\kappa$ is calculated by a weighted amplification of $\kappa_{gr}$, however, the critical curves appear at a larger distance from the origin and their forms are varied to some extent compared to GR. Varying the value of $x_{0}$ from $100$kpc to $400$kpc, we substantially obtain the same findings.

\subsection{Lenses with Multiple Components}
Next, we want to explore gravitational lensing by multiple objects. For this purpose, let us consider a rather simple case and start with two density distributions, $\rho_{1}$ and $\rho_{2}$, following the King profile \eqref{eq:54}. Choosing $r_{c}=200$kpc and $M_{1}+M_{2}=M=10^{15}M_{\odot}$ ($M_{i}$ denotes the total mass of the object located at $\vec{r}_{i}$ inside our volume), we shall place our densities at the following positions inside the grid volume ($r=|\vec{r}|=0$ corresponds to the grid's origin):
\begin{equation}
\vec{r}_{1} =
\begin{pmatrix}
x_{2}\\
0\\
z_{2}
\end{pmatrix},\quad
\vec{r}_{2} = 
-\begin{pmatrix}
x_{2}\\
0\\
z_{2}
\end{pmatrix}.
\label{eq:59}
\end{equation}
Thus, varying the parameters $x_{2}$ and $z_{2}$, we are able to control the relative alignment of our objects along the line of sight, i.e. the $z$-direction, as well as perpendicular to it.

\begin{figure*}[t]

\begin{minipage}[t]{6.7cm}
\begin{center} 
\includegraphics[trim= 5 0 0 0,width=6.2cm]{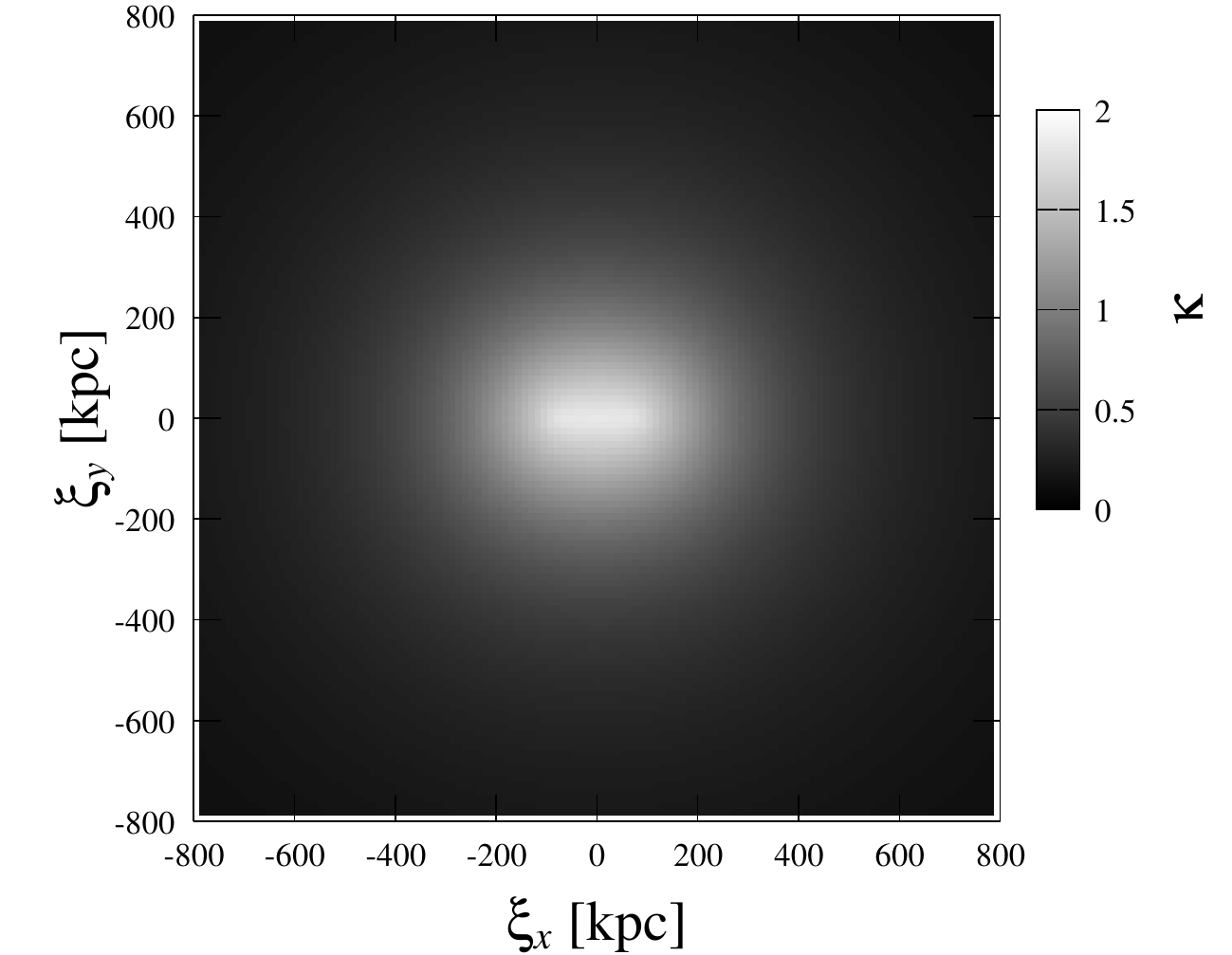}\\[0.3cm]
\includegraphics[trim= 5 0 0 0,width=6.2cm]{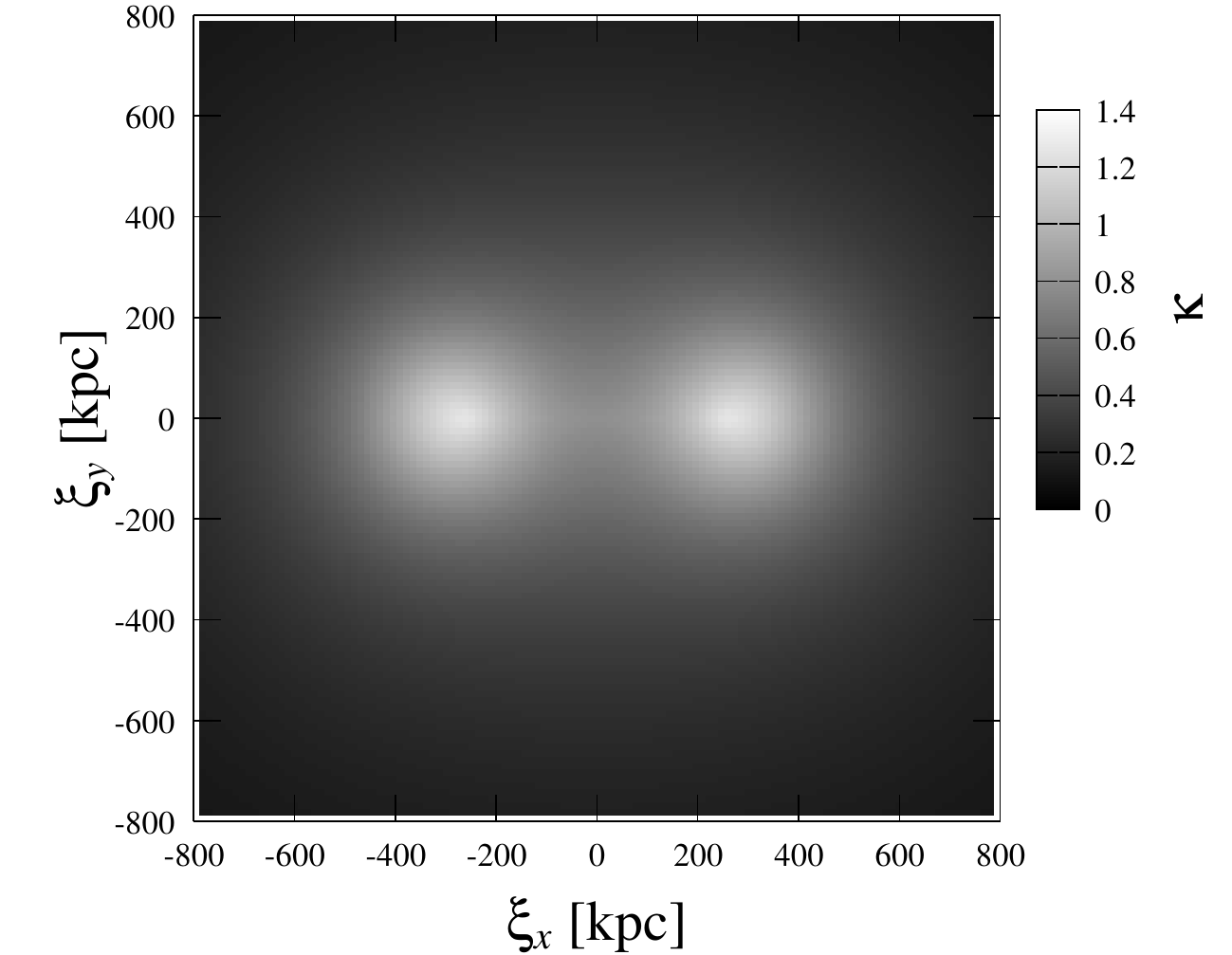}\\[0.3cm]
\includegraphics[trim= 5 0 0 0,width=6.2cm]{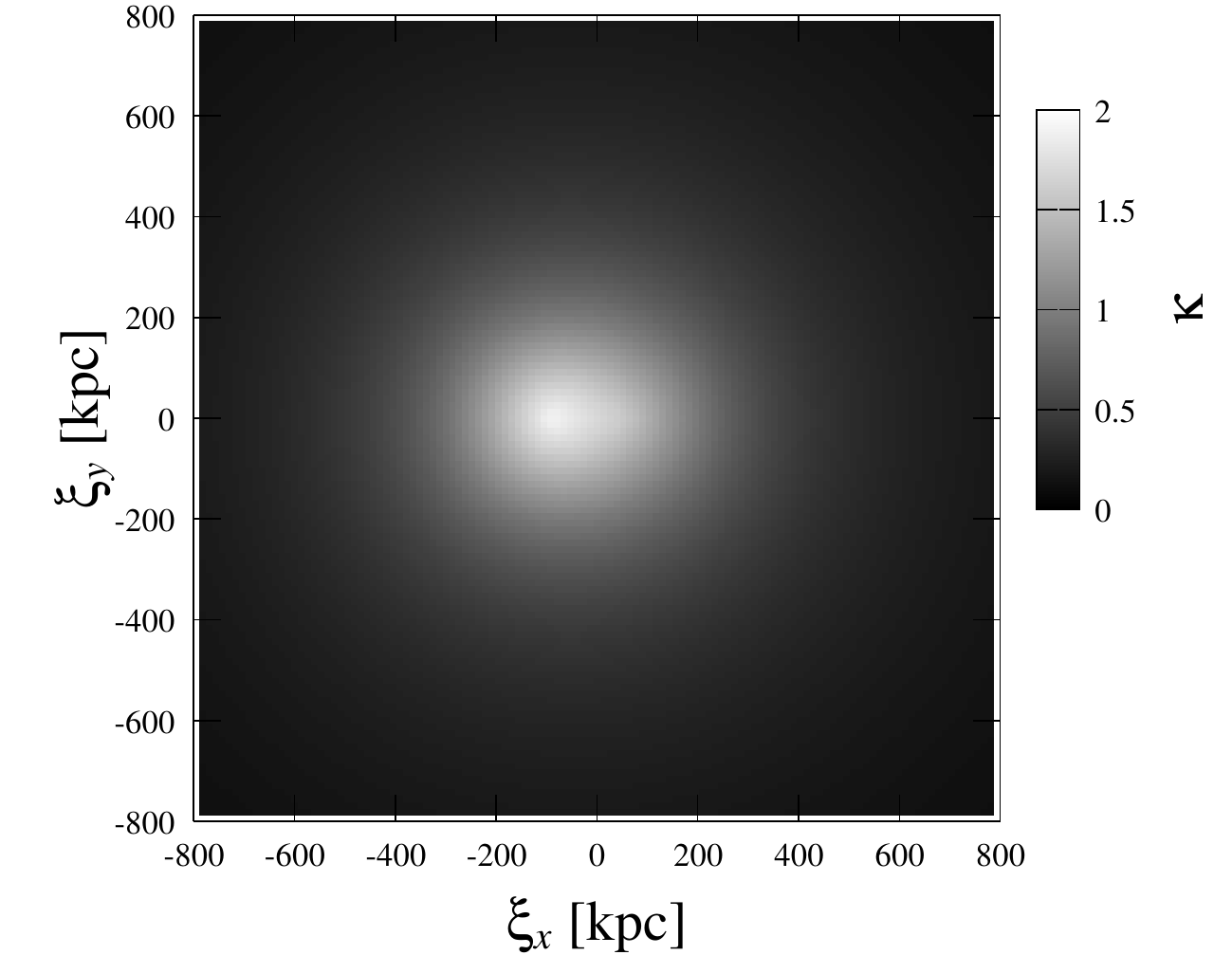}
\end{center}
  \end{minipage}
\hfill
\begin{minipage}[t]{6.7cm}
\begin{center} 
\includegraphics[trim= 50 0 0 0,width=5.5cm]{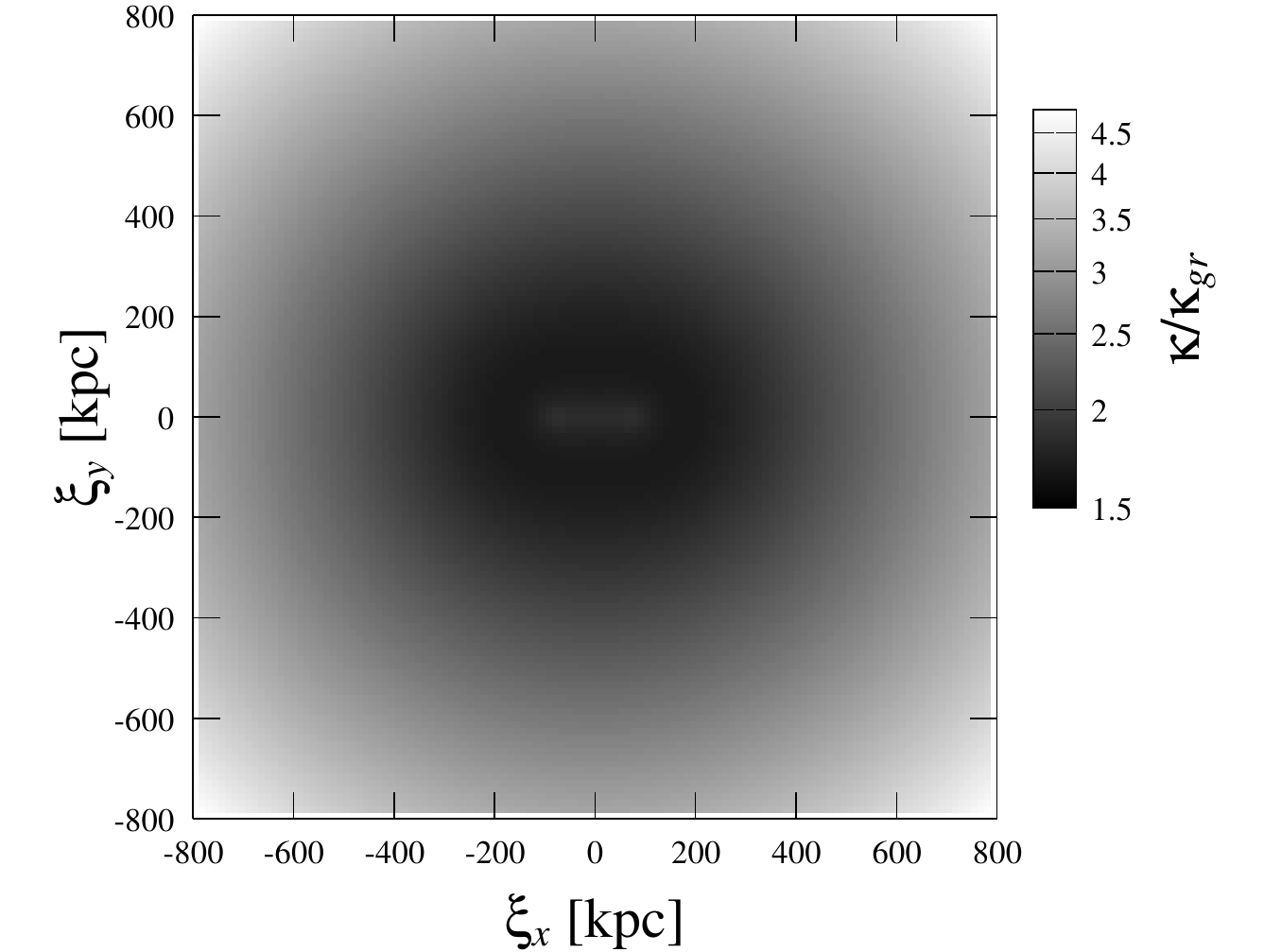}\\[0.4cm]
\includegraphics[trim= 50 0 0 0,width=5.5cm]{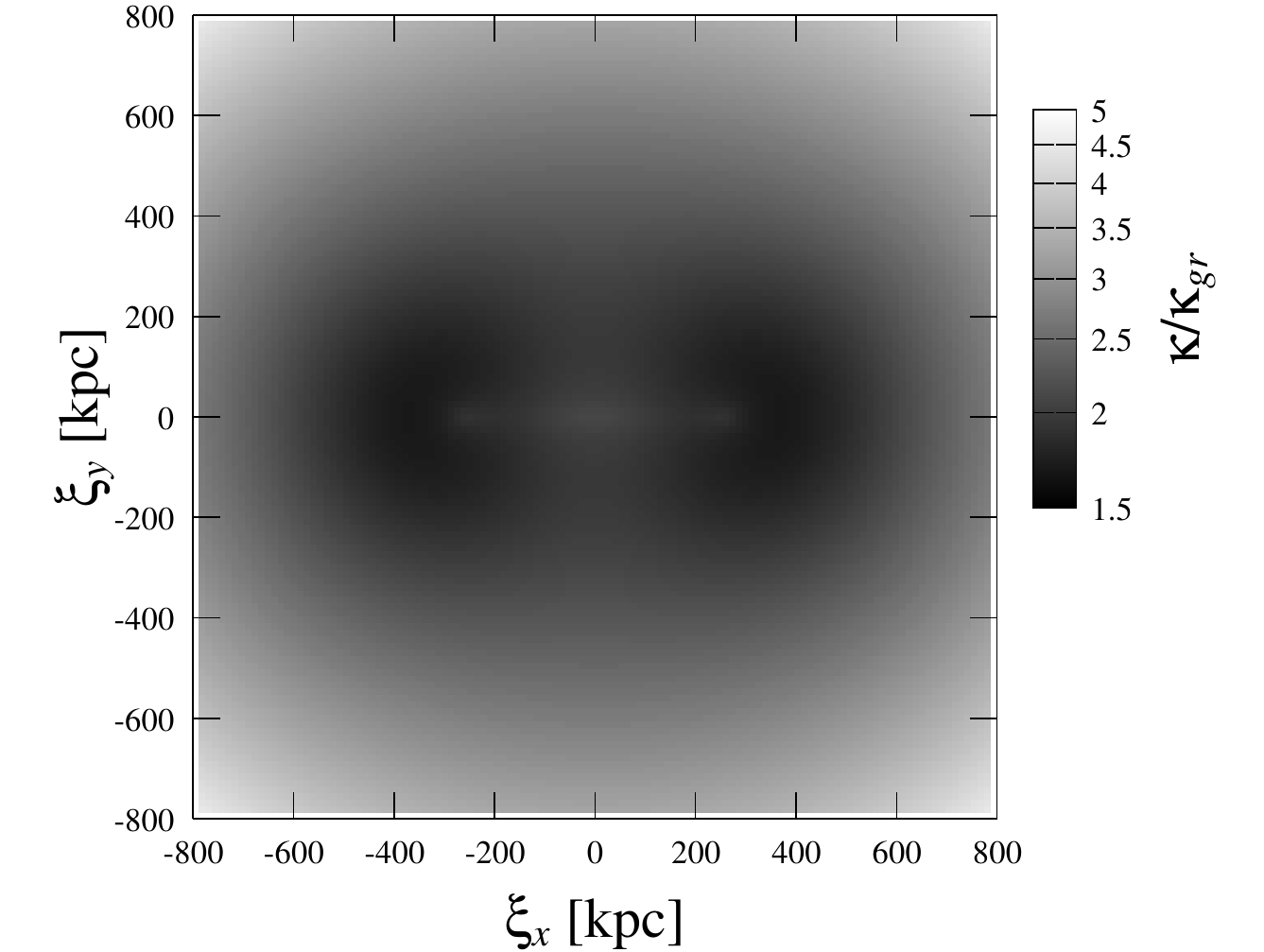}\\[0.4cm]
\includegraphics[trim= 50 0 0 0,width=5.5cm]{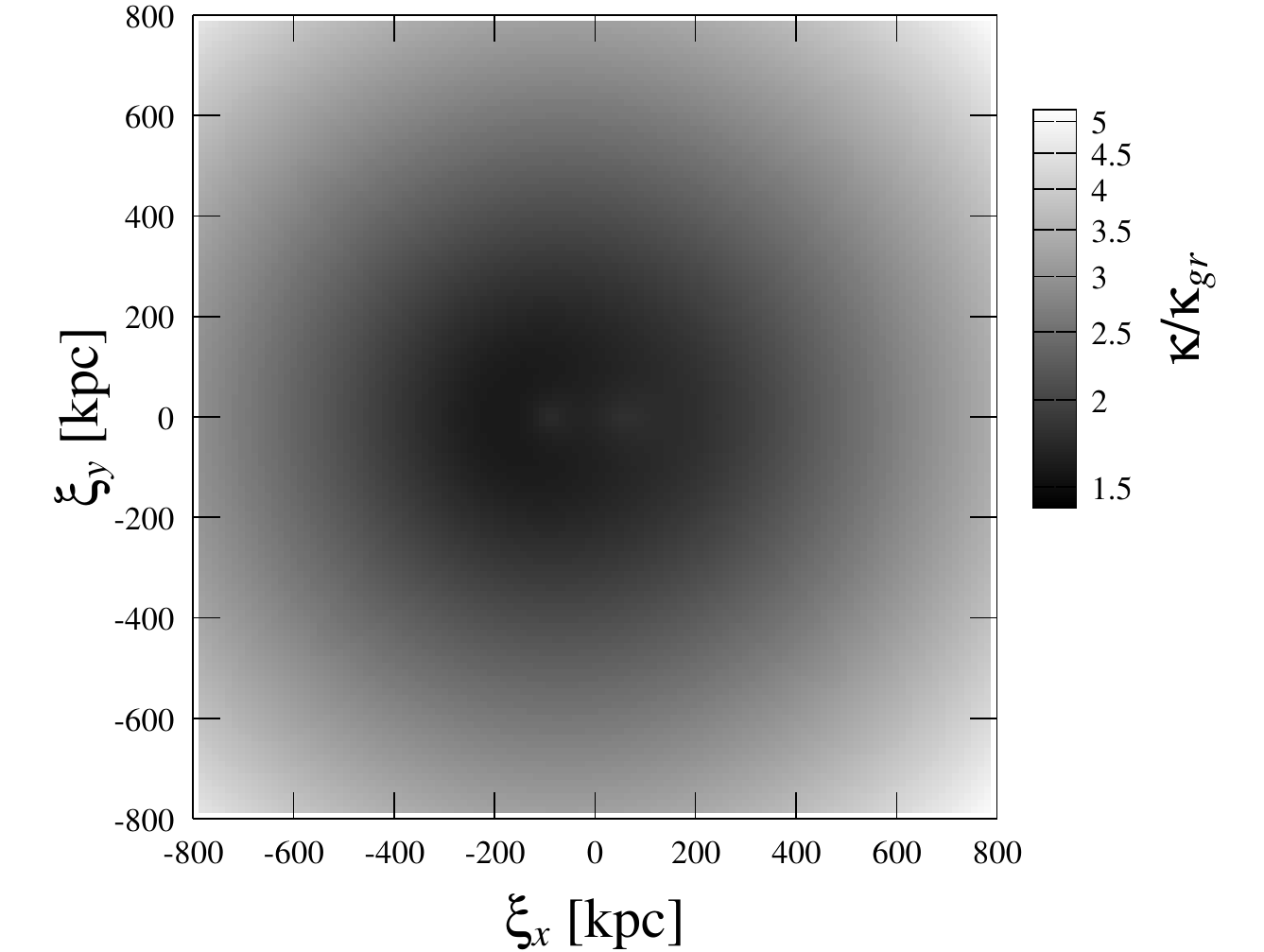}
\end{center}
  \end{minipage}
\hfill
 \begin{minipage}[t]{4.5cm}
\begin{center}
\includegraphics[trim= 4.5cm 1cm 1.5cm 2cm,width=4.5cm]{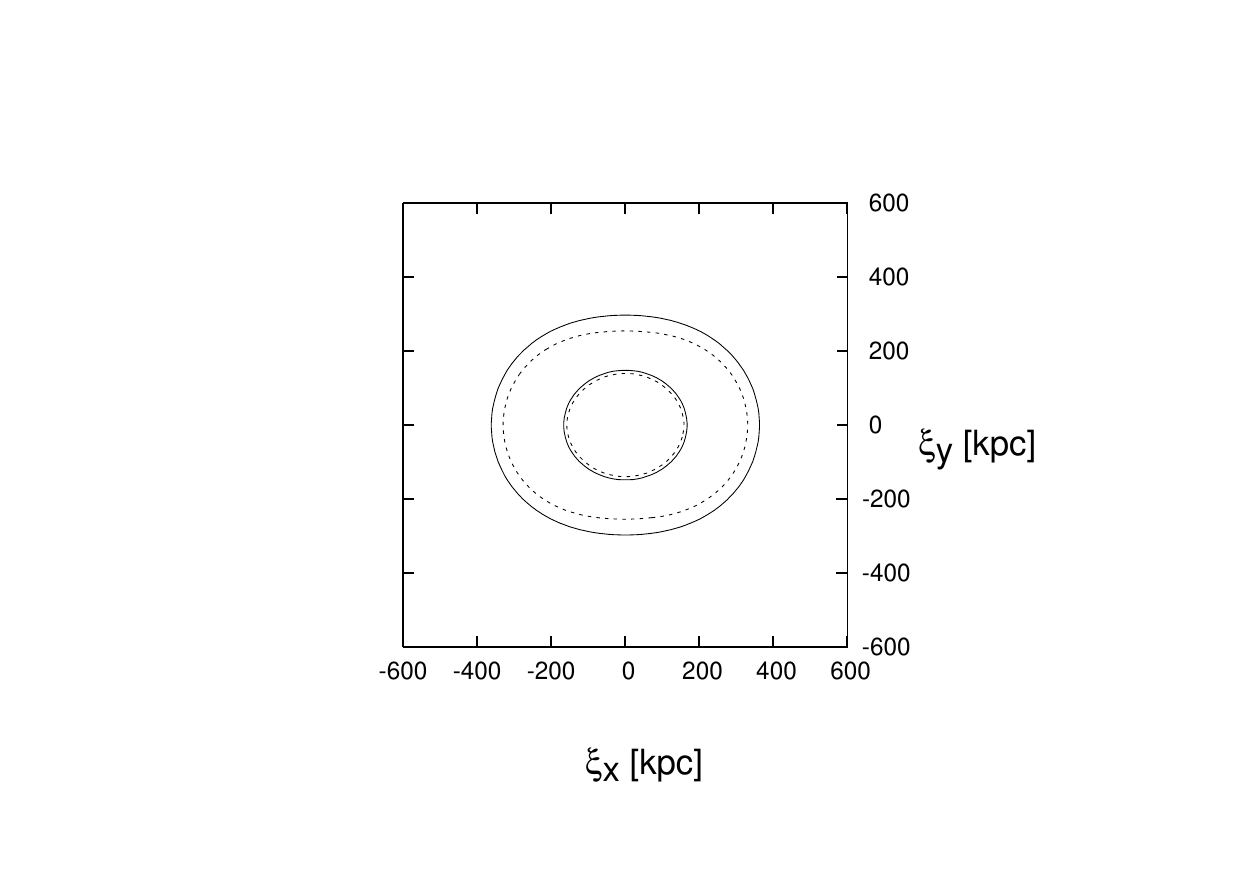}\\[1.2cm]
\includegraphics[trim= 4.5cm 1cm 1.5cm 2cm,width=4.5cm]{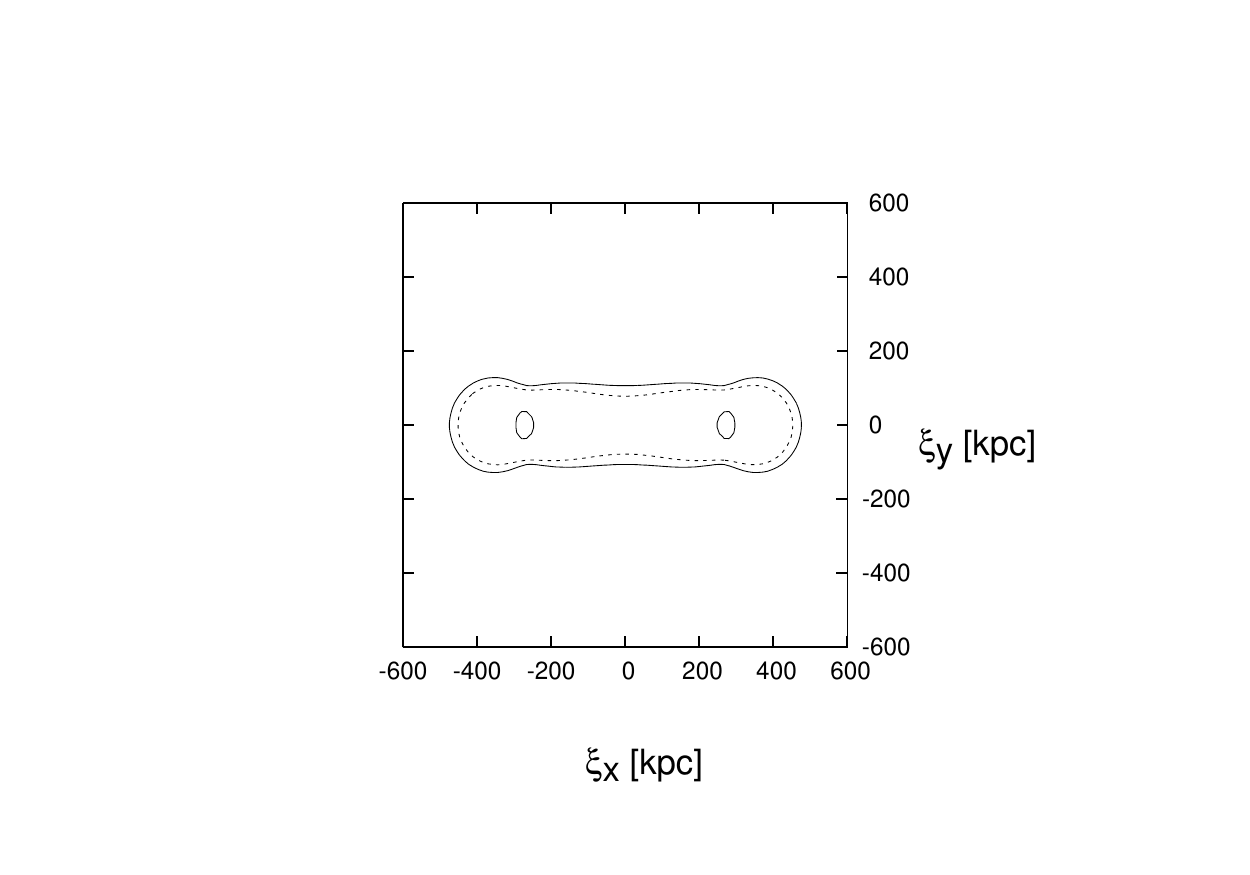}\\[1.2cm]
\includegraphics[trim= 4.5cm 1cm 1.5cm 2cm,width=4.5cm]{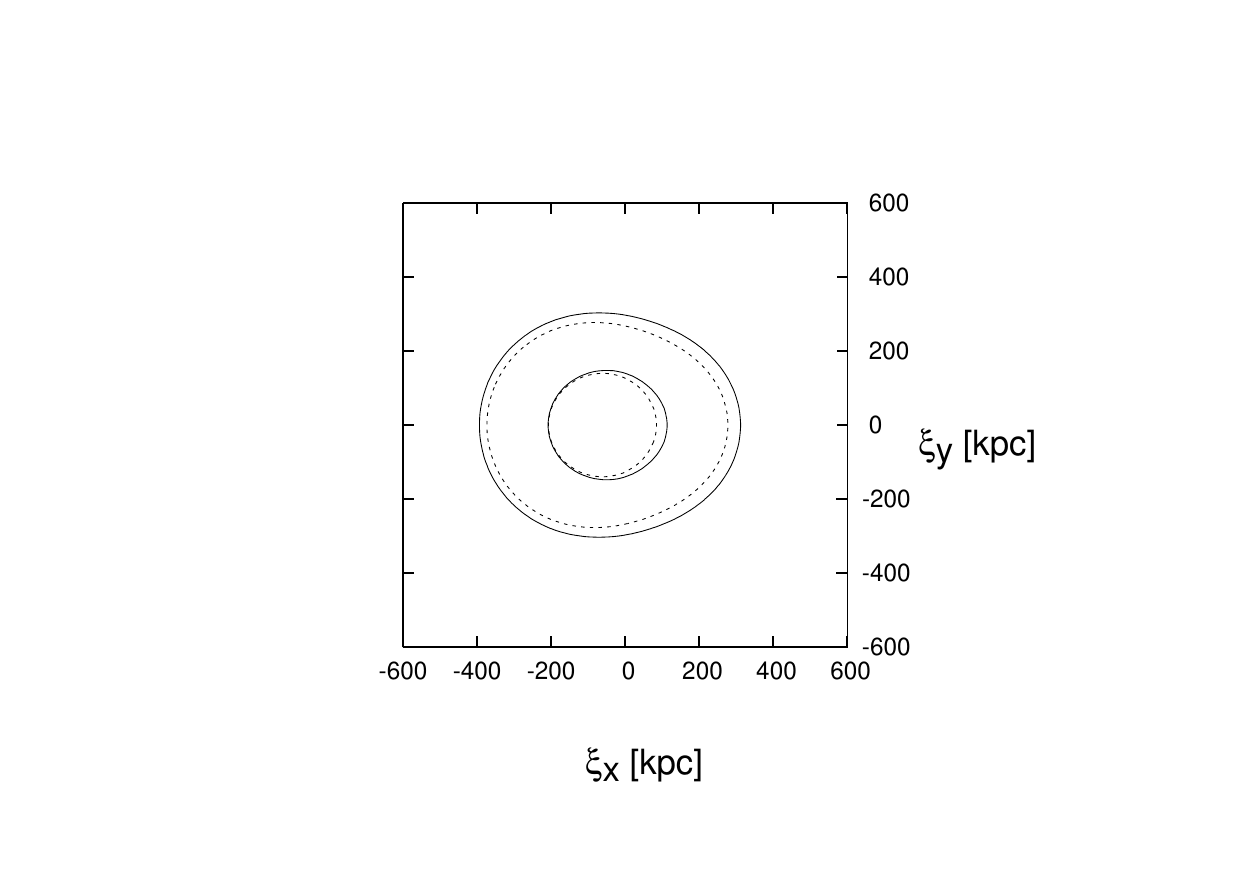}
\end{center}
\end{minipage}

\caption[Lensing properties of our two-bullet system assuming $x_{2}=100$kpc and $M_{1}=M_{2}$]{Lensing properties of our two-bullet system assuming $x_{2}=100$kpc and $M_{1}=M_{2}$ (top panel), $x_{2}=300$kpc and $M_{1}=M_{2}$ (middle panel), and $x_{2}=100$kpc and $3M_{1}=M_{2}$ (bottom panel), respectively: We present the numerical results for both the TeVeS convergence $\kappa$ (left) and the corresponding ratio $\kappa/\kappa_{gr}$ (middle) setting $z_{2}=400$kpc. On the right, the TeVeS critical lines are plotted for $z_{2}=0$ (dashed) and $z_{2}=400$kpc (solid). Note that the radial critical curve for $x_{0}=300$kpc and $z_{2}=0$ does not appear due to the grid's finite resolution.}
\label{bullet1}
\end{figure*}

\subsubsection{Equal Masses}
\label{equal}
As a first approach, we shall assume the total mass $M$ to be evenly distributed on our two bullet-like objects, i.e. $M_{1}=M_{2}$. Varying $x_{2}$ from $100$kpc to $300$kpc, we calculate the convergence maps and critical lines for different alignments along the line of sight, with the results for $z_{2}=400$kpc (and $z_{2}=0$ for the critical curves) presented in Fig. \ref{bullet1} (top and middle panel). Again, we notice that $\kappa$ is amplified such that the symmetry properties of the surface density $\kappa_{gr}$, are virtually conserved, similar to the result found in the last section. Having a look at the central region, we additionally observe that $\kappa$ is increased between the object's positions, which is actually expected since the Newtonian gradient $\vec\nabla\Phi_{N}$ becomes small there. Altogether, as the TeVeS convergence map closely tracks the baryonic matter distribution, and we do not encounter any new surprising TeVeS effects considering our two-bullet system.

Increasing the quantity $z_{2}$, we discover a significant growth of $\kappa$ around the central part, which is in accordance with our previous result from Sec. \ref{thinlens}. Consequently, the corresponding critical lines, shown on the r.h.s of Fig. \ref{bullet1} (top and middle panel), are spatially driven outwards. Please also note that, due to the non-spherical symmetry of our problem, the shape of those curves is slightly changed when varying $z_{2}$.
\begin{table*}
\caption{Component masses and positions for our toy model of the cluster merger $1$E$0657-558$}
\begin{center}
\begin{tabular}{p{3.5cm} c c c c}     
\hline
\noalign{\smallskip}
Component & Position $(x,y,z)$ [kpc]$^{a}$ & Plasma mass $M_{X}$ ($10^{12}M_{\odot}$) & Stellar mass $M_{*}$ ($10^{12}M_{\odot}$) & $M_{total}$ ($10^{12}M_{\odot}$)$^{b}$
\tabularnewline
\noalign{\smallskip}
\hline
\noalign{\smallskip}
Main cluster & $(-350,-50,z_{1})$ & $5.5$ & $0.5$ & $6.0$
\tabularnewline

Main cluster plasma & $(-140,50,z_{2})$  & $6.6$  & $0.2$ & $6.8$
\tabularnewline

Subcluster & $(350,-50,z_{3})$  & $2.7$  & $0.6$ & $3.3$
\tabularnewline

Subcluster plasma & $(200,-10,z_{4})$  & $5.8$  & $0.1$ & $5.9$
\tabularnewline
\noalign{\smallskip}
\hline
\end{tabular}

\end{center}
\label{table1}
\begin{list}{}{}
\item[$^{a}$] For each component, the position perpendicular to the line of sight is approximately determined from the corresponding Magellan and Chandra images.
\item[$^{b}$] Concerning the masses of our toy model components, we use those derived by \cite{bullet}. Note that all masses are averaged within an aperture of $100$kpc radius around the given position.
\end{list}

\end{table*}

\subsubsection{Different Masses}
\label{different}
In analogy to Sec. \ref{equal}, we can perform a similar simulation choosing $3M_{1}=M_{2}$. Assuming $x_{2}=100$kpc, both the calculated convergence map and the critical lines are presented in the bottom panel of Fig. \ref{bullet1} for $z_{2}=400$kpc (and $z_{2}=0$ for the critical curves). As can be seen from the ratio $\kappa/\kappa_{gr}$, the convergence is more strongly amplified in the $\xi_{x}>0$ regime, i.e. the region of lower mass density. Accordingly, the corresponding critical lines are drawn further outwards in that region. As the MONDian influence increases for smaller values of the Newtonian gradient's modulus $|\vec\nabla\Phi_{N}|$, however, this is exactly what one would expect. Choosing other bullet alignments or mass weightings, we basically obtain the same results.
\subsection{Modeling the Bullet Cluster}
Only recently, the cluster merger $1$E$0657-558$, has been announced as a direct empirical proof of the existence of DM \citep{bullet,bullet2} as the weak lensing reconstruction of $\kappa$ shows peaks that are clearly detached from the dominant baryonic components, i.e. the plasma clouds.

Using an analytic model, \cite{tevesfit} have fit this map and derived the corresponding baryonic matter density in MOND-like gravity, concluding that it is not possible to model the merger without assuming an additional invisible mass component located in the central parts of the two clusters. As they have used the weak lensing reconstruction of \cite{bullet}, however, their convergence map does not account for the observed strong lensing features within this system.
\begin{figure*}[p]
\begin{minipage}[t]{1cm}
\begin{center}
\end{center}
  \end{minipage}
\begin{minipage}[t]{6.9cm}
\begin{center}
\includegraphics[width=6.9cm]{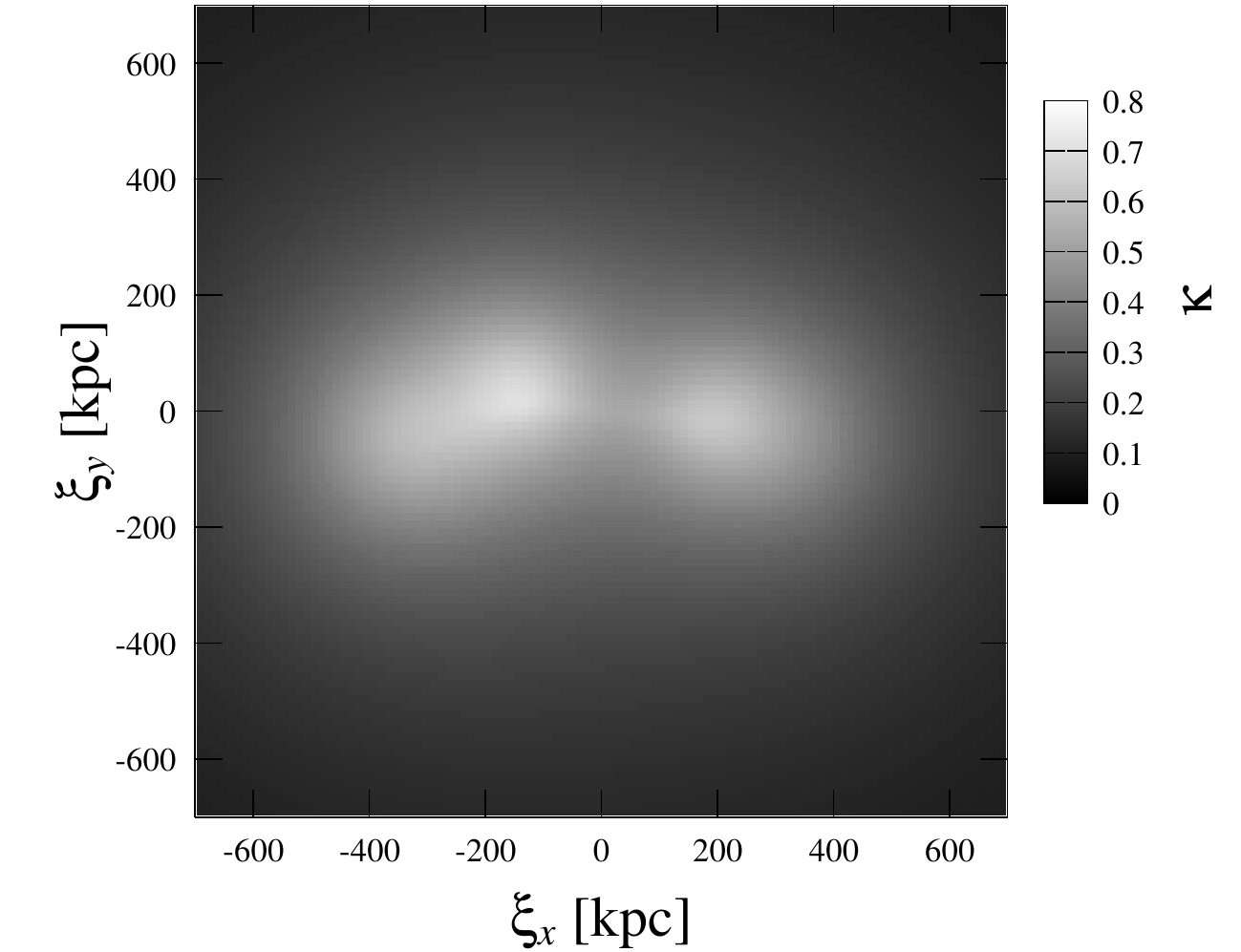}\\[0.5cm]
\includegraphics[width=6.9cm]{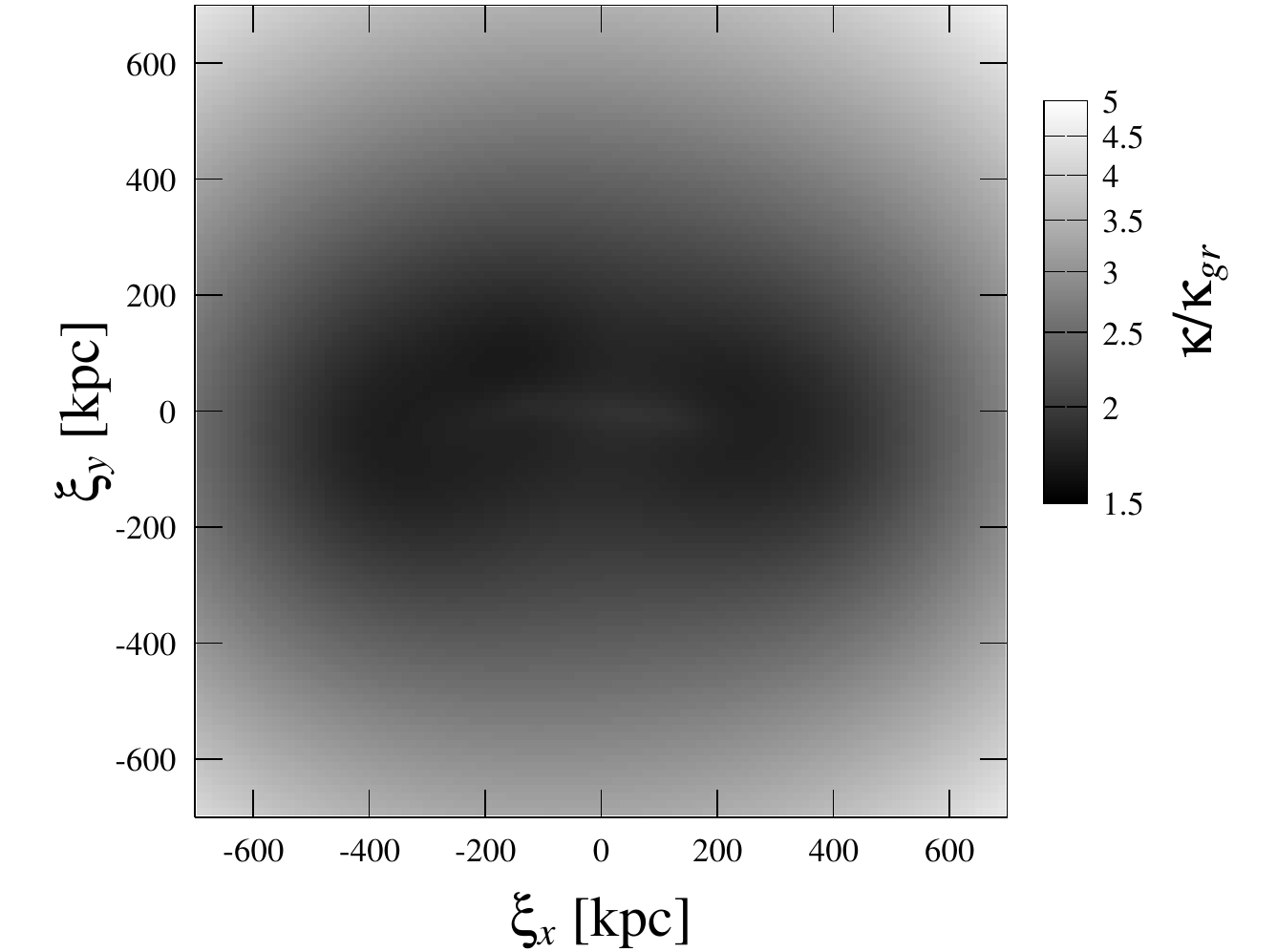}
\end{center}
  \end{minipage}
\hfill
 \begin{minipage}[t]{6.9cm}
\begin{center}
\includegraphics[width=6.9cm]{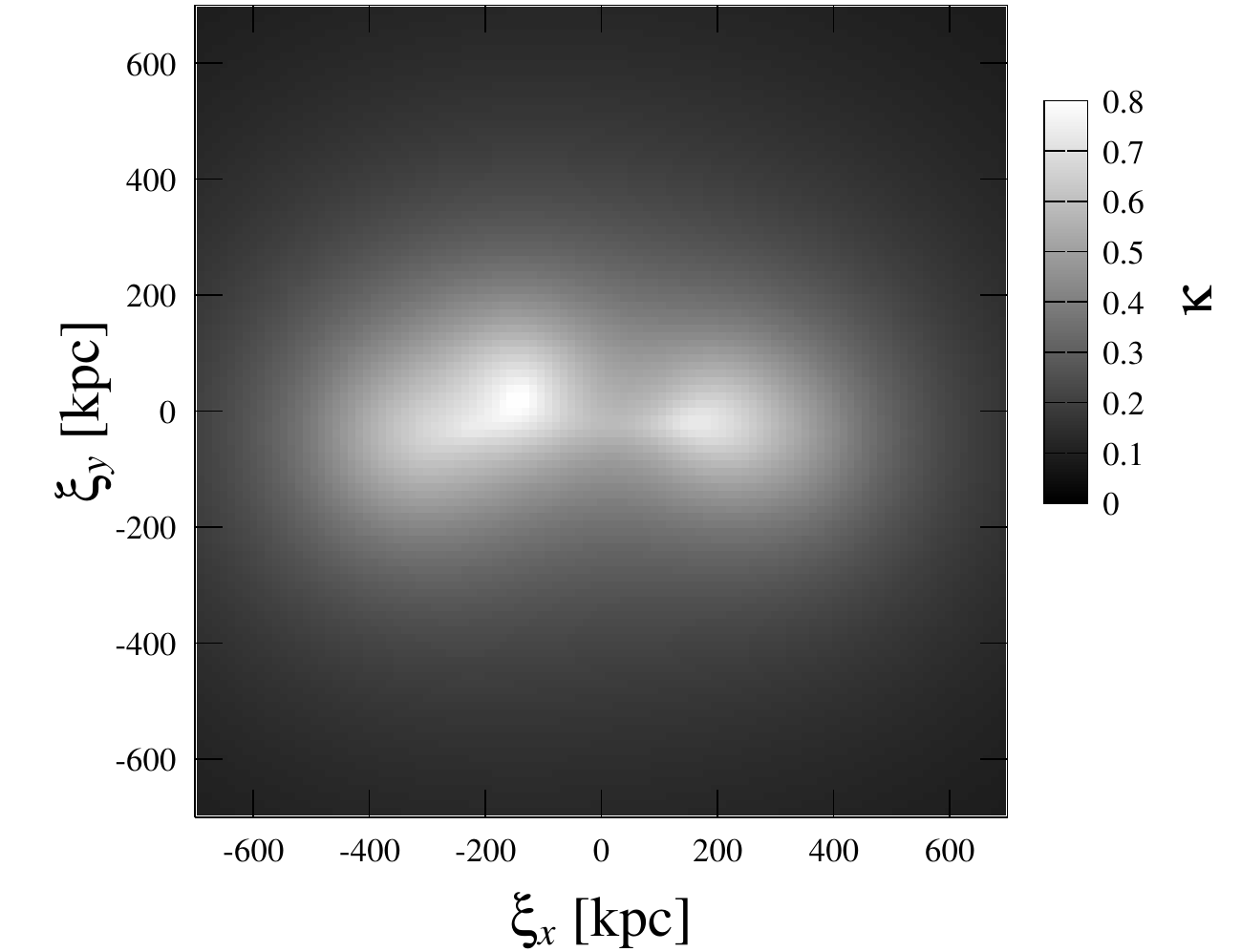}\\[0.5cm]
\includegraphics[width=6.9cm]{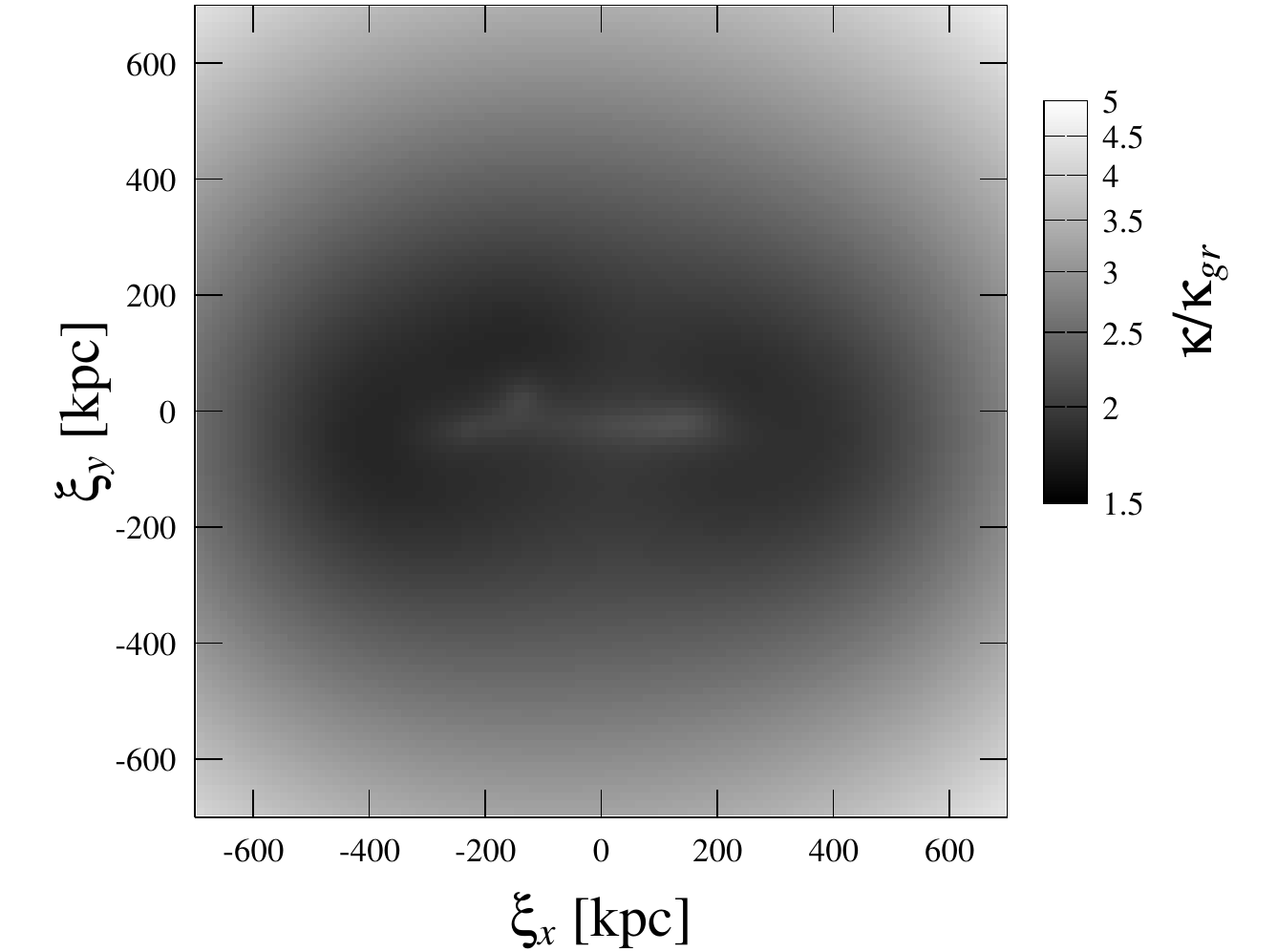}
\end{center}
\end{minipage}
\begin{minipage}[t]{1cm}
\begin{center}
\end{center}
  \end{minipage}
\caption[TeVeS convergence maps for our toy model of the bullet cluster]{TeVeS convergence maps for our toy model of the bullet cluster: Assuming the framework of TeVeS, we present the numerically obtained convergence $\kappa$ (top panel) and the ratio $\kappa/\kappa_{gr}$ (bottom panel) for the parameter sets $A$ (left) and $C$ (right).}
\label{bulletcluster1}
\end{figure*}

\begin{figure*}[p]
\begin{minipage}[t]{1cm}
\begin{center}
\end{center}
  \end{minipage}
\begin{minipage}[t]{6.9cm}
\begin{center}
\includegraphics[width=6.9cm]{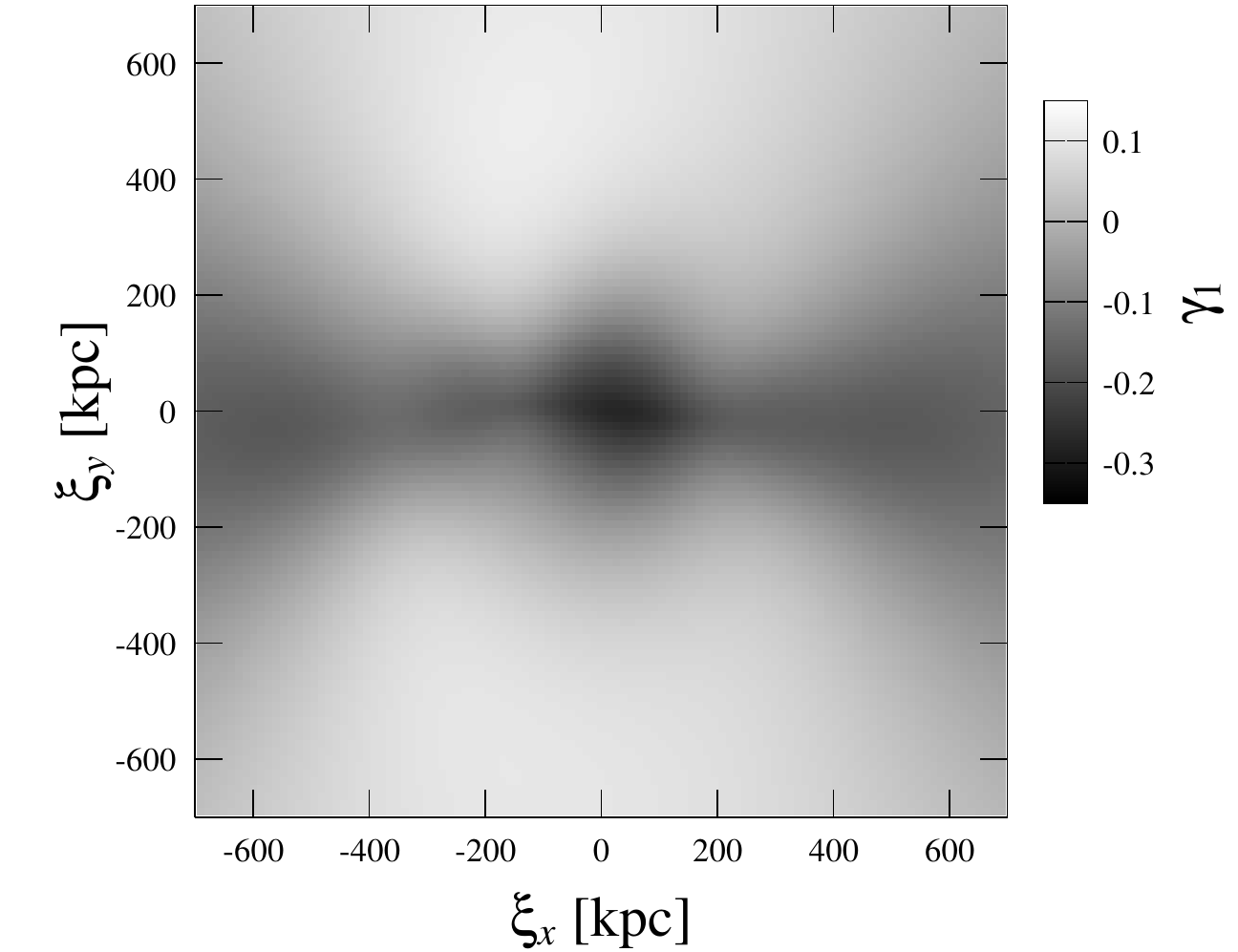}\\[0.5cm]
\includegraphics[width=6.9cm]{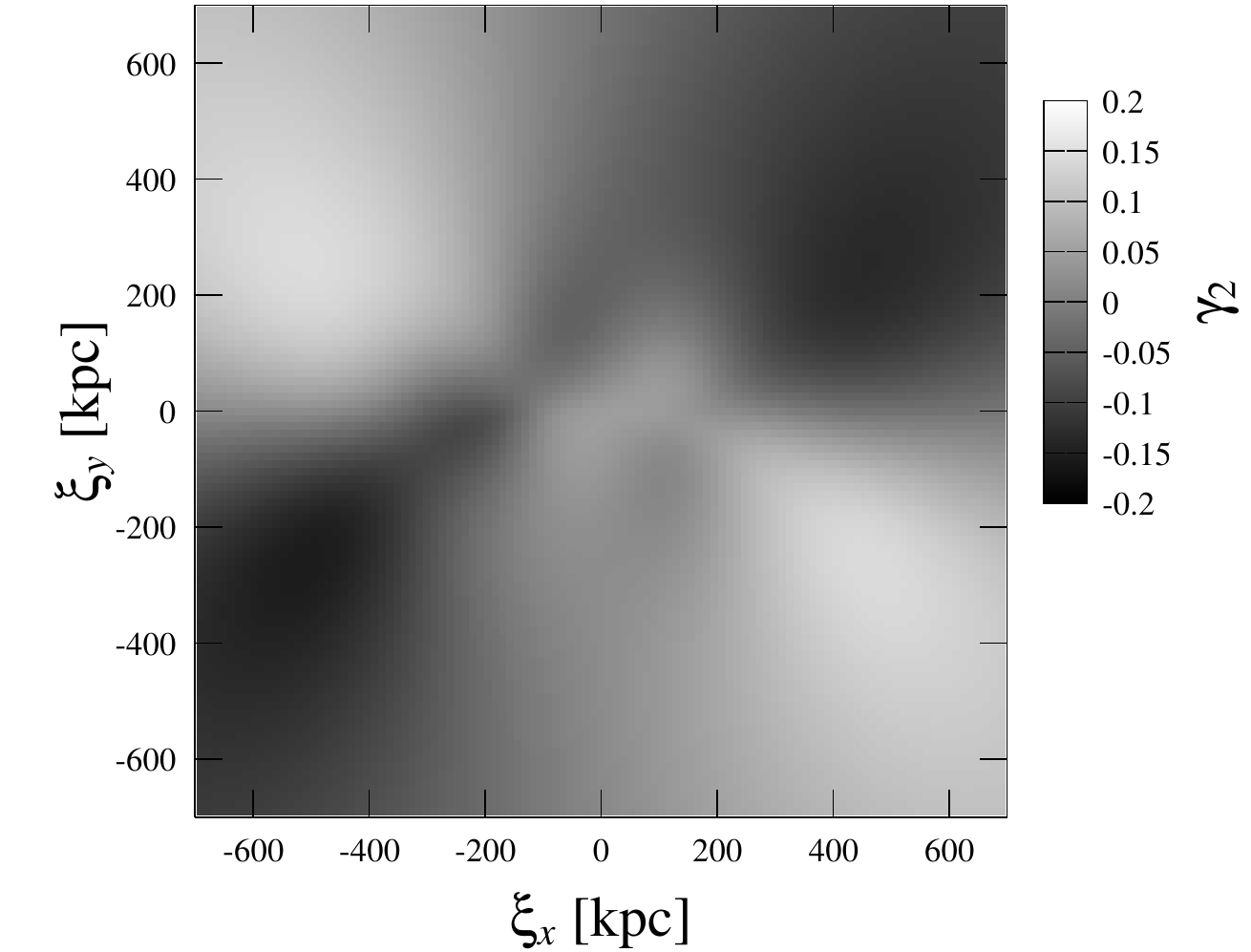}
\end{center}
  \end{minipage}
\hfill
 \begin{minipage}[t]{6.9cm}
\begin{center}
\includegraphics[width=6.9cm]{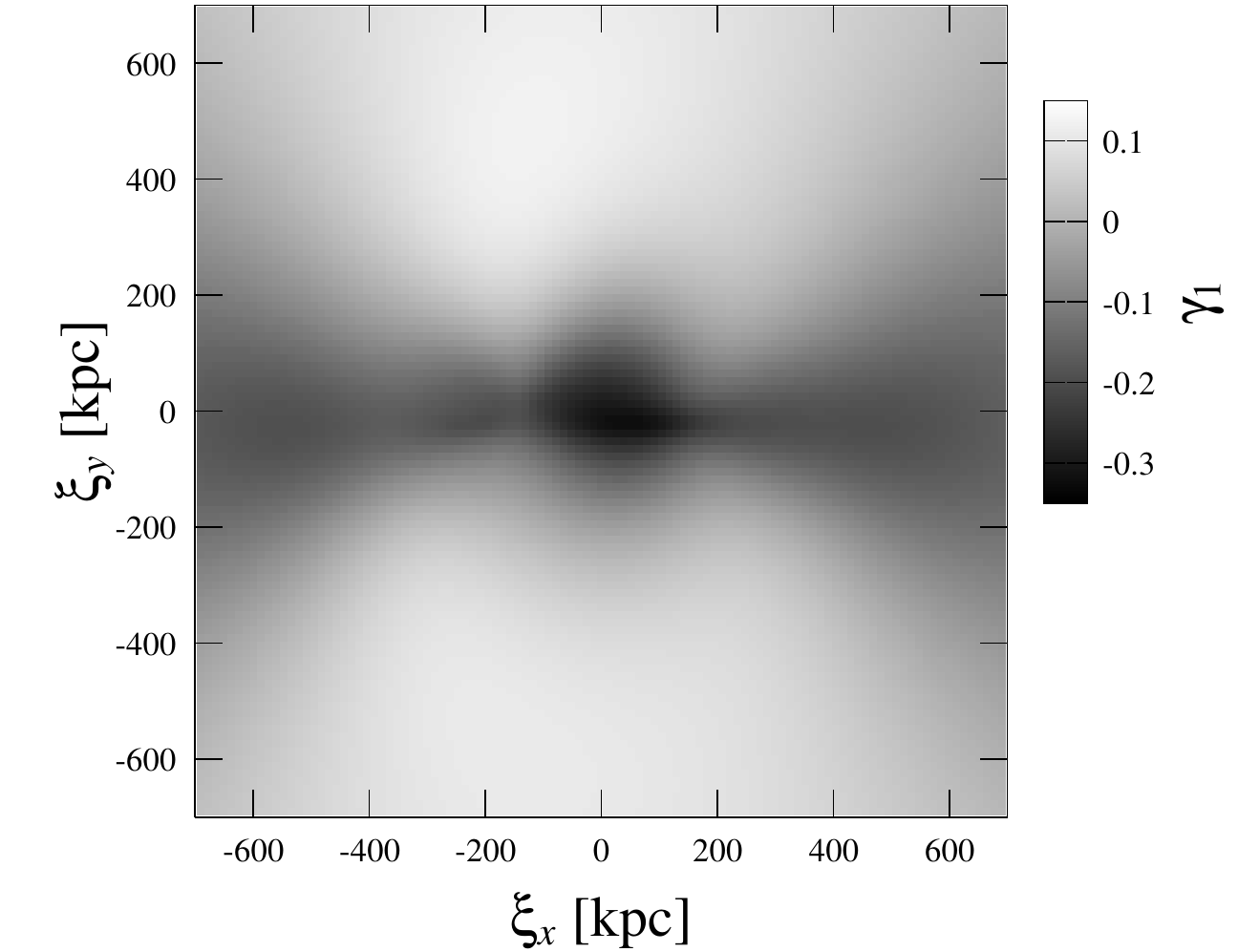}\\[0.5cm]
\includegraphics[width=6.9cm]{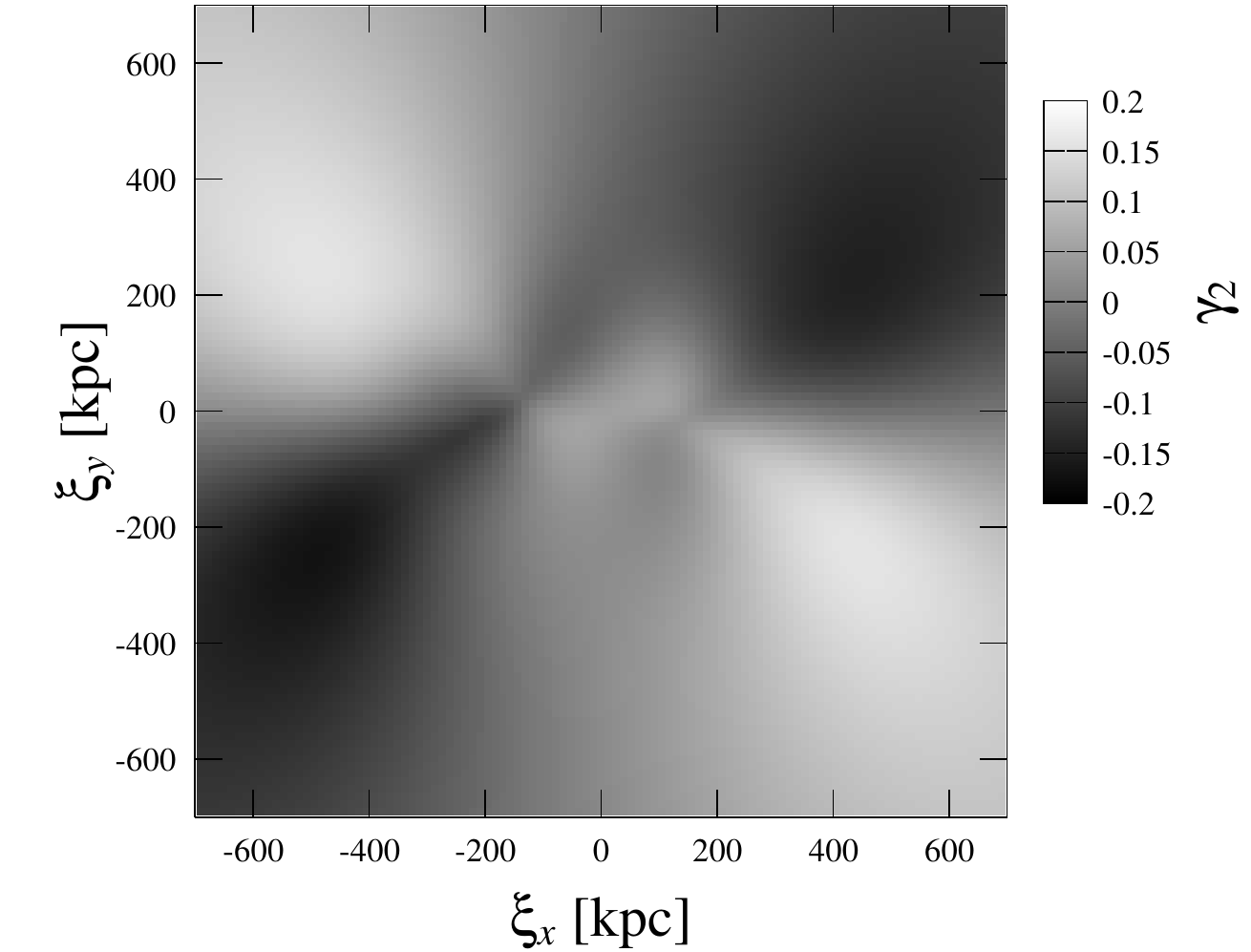}
\end{center}
\end{minipage}
\begin{minipage}[t]{1cm}
\begin{center}
\end{center}
  \end{minipage}
\caption[TeVeS maps of the shear components $\gamma_{1}$,$\gamma_{2}$ for our toy model of the bullet cluster]{TeVeS maps of the shear components $\gamma_{1}$,$\gamma_{2}$ for our toy model of the bullet cluster: Assuming the framework of TeVeS, we present the numerically obtained shear components $\gamma_{1}$ (top panel) and $\gamma_{2}$ (bottom panel) for the parameter sets $A$ (left) and $C$ (right).}
\label{bulletcluster2}
\end{figure*}
In addition, the same convergence map appears to be incompatible with the values of plasma masses independently estimated from Chandra observations, which, as remarked by the authors, may be due to the smoothing scale of the weak lensing reconstruction. To avoid such discrepancies, we shall create a simple toy model of the bullet cluster's baryonic matter density that allows to be treated with our numerical method.

The bullet cluster basically consists of four objects: The main cluster, a slightly smaller subcluster and two plasma clouds appearing in between. For the three-dimensional matter density, we shall model these components using the analytic King profile \eqref{eq:54}, choosing core radii $r_{c}$ of $200$kpc and $150$kpc for the clusters and the plasma clouds, respectively. Concerning the particular masses, we use the values of \cite{bullet} which are derived independently of gravitational lensing. According to the authors, the plasma mass is reconstructed from a multicomponent three-dimensional cluster model fit to the Chandra X-ray image while, assuming a mass-to-light ratio of $M/L_{I}=2$, the stellar mass is calculated from the I-band luminosity of all galaxies equal in brightness or fainter than the component's brightest cluster galaxies (BCG). Together with the approximate positions of the components, the corresponding mass values are presented in Tab. \ref{table1} ($|\vec r|=0$ denotes the grid's center). Please note that all masses are averaged within an aperture of $100$kpc radius around the given position.
For the calculation of the lensing maps, we set the redshift of the bullet cluster, i.e. the lens, to $z_{lens}=0.296$ \citep{bullet} and assume a source redshift of $z_{source}=1$.
\begin{table}
\caption[]{Parameter sets used within the toy model of the cluster merger $1$E$0657-558$}
         \label{sets}
     $$ 
         \begin{array}{p{0.4\linewidth}cccc}
            \hline
            \noalign{\smallskip}
            Parameter set$^{*}$      &  z_{1} & z_{2} & z_{3} & z_{4} \\
            \noalign{\smallskip}
            \hline
            \noalign{\smallskip}
            A & 0 & 0 & 0 & 0 \\
            B & 0 & 300 & 0 & -300 \\
            C & 0 & 500 & 0 & -500 \\
            D & 300 & 100 & -300 & -100             \\
            \noalign{\smallskip}
            \hline
         \end{array}
     $$ 
\begin{list}{}{}
\item[$^{*}$] In our simulations, the above sets are used to specify the component's alignment along the line of sight, i.e. the $z$-direction.
\end{list}
   \end{table}

Since the position of the particular constituents can only be constrained perpendicular to the line of sight, we actually have substantial freedom in selecting their alignment along the $z$-direction.
For our analysis, we choose four different sets of $z_{i}$ which are shown in Tab. \ref{sets}.
Let us briefly discuss the meaning of these choices: Clearly, the parameter set $A$ implies that all components are located in the same plane perpendicular to the line of sight. The choices $B$ and $C$ account for the plasma clouds to be displaced in opposite directions along the $z$-axis, which is a reasonable assumption considering today's view of the bullet cluster to be a post-merger snapshot. Finally, the parameter set $D$ fairly describes the situation of the axis connecting the cluster centers being inclined w.r.t. the $z$-direction.

{ Fig. \ref{bulletcluster1} illustrates both the resulting TeVeS convergence map $\kappa$ and the corresponding ratio $\kappa/\kappa_{gr}$ for the parameter sets A and C listed in Tab. \ref{sets}.
Again, we find that there are TeVeS effects causing additional structure within the central part, and
increasing the constituent's relative displacement along the $z$-axis, we observe these structures growing stronger to some extent.}
Similar to our previously considered lens models, however, the TeVeS effects are not large enough to account for displacements from
the dominant baryonic components. Additionally, we present maps of the TeVeS shear components $\gamma_{1}$ and $\gamma_{2}$ { for the sets A and C} which are shown in Fig. \ref{bulletcluster2}.

Clearly, our result confirms the findings of \cite{tevesfit}. To provide an acceptable explanation of the observations, TeVeS
needs an additional mass component centered at the cluster positions. As has been suggested by others, e.g. \cite{neutrinos}, primordial neutrinos with mass on the order of $2$eV might be able to resolve the problem. { Checking the $y$-values, i.e. the arguments of the free function (cf. Sec. \ref{section22}), near the cluster centers in our simulation, we find that the non-linearity of Eq. \eqref{eq:15} still has a relevant impact on the resulting scalar field. Therefore, it is not possible to isolate neutrino effects as a pure additive contribution to the overall convergence map and to give constraints on the amount and distribution of such neutrinos for a given mass. Current work is trying to find an approximative way of dealing with this issue using our previous numerical results.}

%

\section{Conclusions}

In this work, we have analyzed the effects of gravitational lensing within the framework of TeVeS, focusing on asymmetric systems.

Considering spherically symmetric lenses, we introduced a parameterization of the free function $y(\mu)$ showing that the particular realization of the singularity at $\mu=1$ has a trifling influence on the deflection angle. Furthermore, we concluded that variations of the coupling constant $k$ lead to negligibly small effects as long as $k\lesssim10^{-2}$. Thus, we were able to determine classes of $y(\mu)$ that nearly produce the same deflection angle.

Choosing a single form of the free function, we succeeded in building a fast Fourier-based solver for scalar potential $\phi$ which could be applied to a set of different non-spherical lens types. Concerning variations on rather small scales, we noticed a strong dependence of the lensing properties on the lens's extent along the line of sight, with a significant impact on the critical curves. Additionally, every simulated TeVeS convergence map showed a strong resemblance with the dominant baryonic mass components, other effects, being capable of counteracting this trend, turned out to be very small. To study a more complex lens system, we finally created a toy model of the bullet cluster's baryonic matter density. The outcome of our simulation clearly confirms the results of \cite{tevesfit} as it is not possible to explain the observed weak lensing map without assuming an additional dark mass component in both cluster centers.

Future work will address even more complex lens systems like, for example, the galaxy cluster Abell $2390$ with its straight arc. Increasing the lens's level of substructure, effects that account for loosing track of the projected matter density could become more important, thus influencing the ability of TeVeS to model such a lens. In addition, one can use these models to check if TeVeS is consistent with the assumption of massive neutrinos in galaxy clusters.



\bibliography{aa_old/ref}



\end{document}